\documentclass{ws-ijmpe}
\usepackage[super,compress]{cite}

\def\etal{{\it et al.}}
\def\go{\rightarrow  }
\def\be{\begin{equation}}
\def\ee{\end{equation}}
\def\br{\begin{eqnarray}}
\def\er{\end{eqnarray}}
\def\brn{\begin{eqnarray*}}
\def\ern{\end{eqnarray*}}
\def\rf#1{{(\ref{#1})}}

\def\I {{{\cal I}}}
\def\M {{{\cal M}}}
\def\O {{{\cal O}}}

\def\T {{{\cal T}}}

\def\bit{\begin{itemize}}
\def\eit{\end{itemize}}
\def\Ket#1{||#1 \rangle}
\def\Bra#1{\langle #1||}
\def\lsim{\:\raisebox{-0.5ex}{$\stackrel{\textstyle<}{\sim}$}\:}
\def\gsim{\:\raisebox{-0.5ex}{$\stackrel{\textstyle>}{\sim}$}\:}
\def\ie{{\it i.e., }}

\def\nn{\nonumber }

\def\ket#1{|#1 \rangle}
\def\rf#1{{(\ref{#1})}}

\def\sixj#1#2#3#4#5#6{\left\{\negthinspace\begin{array}{ccc}
#1&#2&#3\\#4&#5&#6\end{array}\right\}}
\def\ninj#1#2#3#4#5#6#7#8#9{\left\{\negthinspace\begin{array}{ccc}
#1&#2&#3\\#4&#5&#6\\#7&#8&#9\end{array}\right\}}
\def\go{\rightarrow  }
\def\sqi{\frac{1}{\sqrt{2}}}
\newcommand{\Mass}{\mathrm{M}}

\def\fot{\frac{1}{2}}

\def\rf#1{{(\ref{#1})}}

\def\ket#1{|#1 \rangle}
\def\Ket#1{||#1 \rangle}
\def\Bra#1{\langle #1||}

\def\be{\begin{equation}}
\def\ee{\end{equation}}
\def\ber{\begin{eqnarray}}
\def\eer{\end{eqnarray}}
\def\etc{{\it etc}}

\def\mblambda{\mbox{\boldmath$\lambda$}}
\def\mbl{\mbox{\boldmath$l$}}
\def\mbL{\mbox{\boldmath$L$}}

\def\bnu{\begin{enumerate}}
\def\enu{\end{enumerate}}
\begin{document}

\markboth{Franjo Krmpoti\'c and Cl\'audio De Conti}{Nonmesonic Weak Decay Dynamics from Proton  Spectra of $\Lambda$-hypernuclei}

\catchline{}{}{}{}{}

\title{NONMESONIC WEAK DECAY DYNAMICS FROM PROTON SPECTRA OF $\Lambda$-HYPERNUCLEI }

\author{\footnotesize FRANJO KRMPOTI\'C }

\address{Instituto de F\'{\i}sica La Plata, Universidad Nacional de La Plata\\
1900 La Plata, Argentina}

\address{Instituto de F\'isica Te\'orica, UNESP - Univ. Estadual Paulista, \\
S\~ao Paulo, S\~ao Paulo 01140-070, Brasil \\
krmpotic@fisica.unlp.edu.ar}


\author{CL\'AUDIO DE CONTI}

\address{Campus Experimental de Itapeva, UNESP - Univ Estadual Paulista, \\
Itapeva, S\~ao Paulo 18409-110, Brasil \\
conti@itapeva.unesp.br}

\maketitle


\begin{abstract}
A novel comparison between the data and the theory is proposed for
the nonmesonic (NM) weak decay of hypernuclei. Instead of
confronting the primary decay rates, as  is usually done, we focus
attention on the effective decay rates that are straightforwardly
related with the number of emitted particles. Proton kinetic
energy spectra of ${\mathrm{^{5}_{\Lambda}He}}$,
${\mathrm{^{7}_{\Lambda}Li}}$, ${\mathrm{^{9}_{\Lambda}Be}}$,
${\mathrm{^{11}_{\Lambda}B}}$, ${\mathrm{^{12}_{\Lambda}C}}$,
${\mathrm{^{13}_{\Lambda}C}}$, ${\mathrm{^{15}_{\Lambda}N}}$ and
${\mathrm{^{16}_{\Lambda}O}}$, measured by FINUDA, are evaluated
theoretically. The Independent Particle Shell Model (IPSM) is used
as the nuclear structure framework, while the dynamics is
described by  the  One-Meson-Exchange (OME) potential. Only for
the ${\mathrm{^{5}_{\Lambda}He}}$, ${\mathrm{^{7}_{\Lambda}Li}}$,
and ${\mathrm{^{12}_{\Lambda}C}}$ hypernuclei is it possible to
make a comparison with the data, since for the rest there is no
published  experimental information on  number of produced
hypernuclei. Considering solely the one-nucleon-induced ($1N$-NM)
decay channel, the theory reproduces correctly the shapes of all
three spectra at medium and high energies ($E_p \gsim 40$ MeV).  
Yet, it greatly overestimates their magnitudes, as well as the
corresponding transition rates
 when the full  OME ($\pi+K+ \eta+\rho+\omega+K^*$) model is used. The agreement is much improved when
only the $\pi+K$ mesons with soft  dipole cutoff parameters
participate  in the decay process. We find that the IPSM is a fair
first order approximation to disentangle  the dynamics of the
$1N$-NM decay, the knowledge of which is indispensable to inquire
about  the baryon-baryon strangeness-flipping  interaction. It is
shown that the IPSM provides very useful insights regarding the
determination  the $2N$-NM decay rate. In a new analysis of the
FINUDA data, we derive two results for this quantity with one of
them close to that obtained previously.
\end{abstract}


\ccode{PACS numbers: 21.80.+a, 13.75.Ev, 21.60.-n}


\section{Introduction}
\label{Sec1}

The weak decay rate of a $\Lambda$ hypernucleus can be expressed as~\cite{Al02}
\ber
\Gamma_W = \Gamma_M +\Gamma_{NM},
\label{1}\eer
where $\Gamma_M$ is the decay  rate for the mesonic (M) decay $\Lambda
\rightarrow \pi N$,  and  $\Gamma_{NM}$ is the rate for the
nonmesonic (NM) decay, which can be induced either by one bound
nucleon ($1N$), $\Gamma_1^0(\Lambda N  \rightarrow nN)$,
or by two bound nucleons ($2N$), $\Gamma_2^0(\Lambda NN \rightarrow
nNN)$, or even more bound nucleons \ie
\ber
 \Gamma_{NM}=\Gamma_1^{0}+\Gamma_2^0+\cdots;~~
\hspace{0.1cm}\Gamma_{1}^0=\Gamma_p^0+\Gamma_n^0,
\hspace{0.1cm}\Gamma_{2}^0=\Gamma_{nn}^0+\Gamma_{np}^0+\Gamma_{pp}^0.
 \label{2}\eer
With the symbol $\cdots$ we indicate that
additional processes, such as those induced by three nucleons, can contribute also.
We use the superscript $0$ to distinguish between   the primary (bare) decay
rates, and the effective decay rates
\ber
\Gamma_{p}&=&\Gamma_p^{0,FSI}
+\Gamma_n^{0,FSI}+\Gamma_{np}^{0,FSI} +2\Gamma_{pp}^{0,FSI}+\cdots,
\nn\\ \Gamma_{n}&=&\Gamma_p^{0,FSI}+2\Gamma_n^{0,FSI}+2\Gamma_{np}^{0,FSI}+\Gamma_{pp}^{0,FSI}
+3\Gamma_{nn}^{0,FSI}+\cdots,
\nn\\\Gamma_{np}&=&\Gamma_p^{0,FSI}+2\Gamma_{np}^{0,FSI}+2\Gamma_{pp}^{0,FSI}+\cdots,
\nn\\ \Gamma_{nn}&=&\Gamma_n^{0,FSI}+3\Gamma_{nn}^{0,FSI}+\Gamma_{np}^{0,FSI}+\cdots.
 \label{3}\eer
which are affected by  final state interactions (FSIs) and  are directly related to  the numbers  of measured single-nucleons
${\rm N}_N$, and  two-particle coincidences ${\rm N}_{nN}$,   as
\ber
\Gamma_{N}=\frac{\Gamma_W}{{\rm N}_W}{\rm N}_N,\hspace{0.5cm}
\Gamma_{nN}=\frac{\Gamma_W}{{\rm N}_W}{\rm N}_{nN},
 \label{4} \eer
where the number of produced hypernuclei  ${\rm N}_W$ and the corresponding decay rate $\Gamma_W$
are experimentally measured quantities. The primary nucleons, in propagating within
the nuclear environment, interact with the surrounding nucleons
 representing a complicated many-body problem
  generically designated as FSIs. In Eq. \rf{3} we have considered only the dominant primary decays that are later perturbed by FSIs, and this is the meaning of $+\cdots$.
  
\begin{figure}[th]
\centerline{\psfig{file=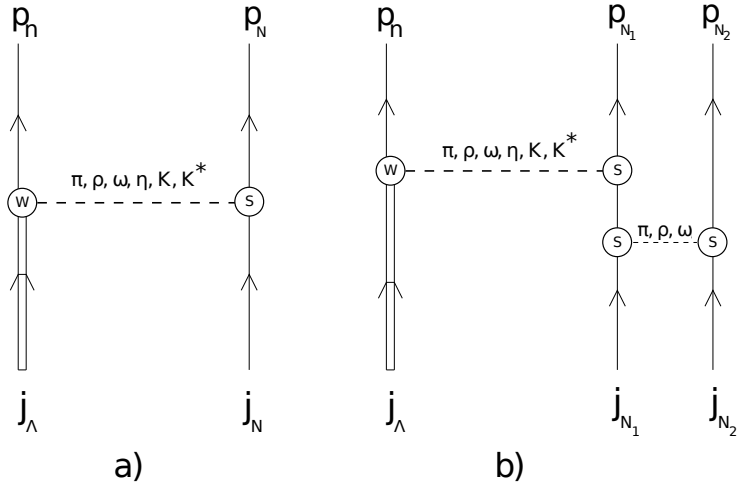,width=10cm}}
\caption{\label{Fig1} Schematic representation of a)  one-nucleon,
and  b) two-nucleon induced decays in $\Lambda$-hypernuclei when
described by the interplay of weak ($W$) and  strong ($S$)
interactions through the exchanges of nonstrange-mesons
$\pi,\rho,\omega$, and $\eta$, and strange-mesons $K$, and $K^*$.
The $W$ and $S$ vertices are exchanged for strange mesons. }
\end{figure}

The  schematic representation of the two decay  channels, when the
pertinent dynamics is described by one-meson-exchange (OME)
potentials, is shown in Fig. \ref{Fig1}.
 This  is the most frequently used model for handling the NM-decay,
 including usually
  the exchanges of  nonstrange-mesons $\pi,\rho,\omega$, and $\eta$, and strange-mesons $K$, and $K^*$.
It is based on the original idea of
 Yukawa that the $NN$  interaction at long distance is due to the one-pion-exchange (OPE), with
the dominant role  played  by the
exchange of  pion and kaon mesons.

The OPE potential was verified quantitatively by the Nijmegen
partial wave analysis of  NN scattering in the elastic
region~\cite{St94}, \ie at distances larger than the minimal de
Broglie wavelength $1/\sqrt{m_\pi\Mass}\sim 0.5$ fm corresponding
to the pion production threshold. The verification of other meson
exchanges is less straightforward, and the uncertainties in the
baryon-baryon-meson (BBM) coupling constants  could be sizeable
since they are not constrained by experiments. To derive them in
the strong sector (S vertices in Fig. \ref{Fig1}), the
$SU(3)_f$ (flavor) symmetry  is  utilized.  In the weak sector (W vertex in Fig.
\ref{Fig1}), the BBM parity-violating couplings are obtained from
the $SU(6)_W$ (weak) symmetry, while the parity conserving couplings
are derived  from  a pole-model with only baryon pole
resonances~\cite{Du96,Pa97}.

The BBM vertex functions  also involve uncertainties in the dipole cutoff parameters $\Lambda_M$
which,  being off-shell quantities, can not be determined experimentally.
 We only know that, to have a physical meaning, they have to be of   hadronic scale ($\sim 1$ GeV).
   For instance, in different calculations,  the   $\Lambda_\pi$  for the off-shell
pion varies from $0.7$ to $1.3$ GeV~\cite{Du96,Pa97,Boc99,Nak12,Ryc94,Bau09,Kr14}.
In particular, the pion cutoffs of $1.2$ GeV, and  $0.8$ GeV were used to describe
the electromagnetically induced two-nucleon emission processes $(\gamma,pn)$ and $(\gamma,pp)$~\cite{Ryc94}, which are quite similar to the $1N$-NM
decays $(\Lambda p,np)$, and $(\Lambda n,nn)$.
We use here the $\Lambda_M$'s
from Ref.~\cite{Pa97}, and also those proposed in Ref.~\cite{Bau09,Kr14}
to account for the  transition rates $\Gamma_p$
 and  $\Gamma_n$ in the $s$-shell hypernuclei.
\footnote{  The dependence of the NMWD transition rates on the values of  $\Lambda_M$, and therefore
    on the BBM vertex functions, is
   thoroughly discussed in  Table 1 and Fig. 1 of Ref.~\cite{Bau09}. }

The short-range correlations (SRCs) between the emitted nucleons  $nN$, and $nNN$ can also affect
significantly the BBM couplings.
 Parre\~{n}o and  Ramos have shown that they can diminish the value of $\Gamma^0_1$ by more than
 a factor of two~\cite{Pa02}.
 Nothing has been stated so far regarding the effect of the SRCs on the $2N$-NM decay.
The theoretical scene becomes even more complex when effects of the quark degrees of freedom~\cite{Sa00,Sa02},
the $2\pi$-exchanges~\cite{Pa04,Ch07,It08,It10}, and the axial-vector $a_1$-meson exchange~\cite{It08,It10} are considered.

The M and  $1N$-NM decays have been observed experimentally in
the pioneering measurement performed  more than 50 years ago by
 Ruderman and R. Karplus~\cite{Ru56}. Conversely,
 the experimental observation of the $2N$-NM decay, which was predicted
 by Alberico \etal~\cite{Al91} in 1991
 (see also Ref.~\cite{Ra94}), has been reported only in recent years
 at KEK~\cite{Ki09}, and  at FINUDA~\cite{Ag08,Ag09,Ag10}.
Both groups announced  a  branching ratio
$\Gamma_{2}^0/\Gamma_{NM}\sim 25-30~\%$. The first group obtained
this  result from  the single and double coincidence nucleon
spectra in ${\mathrm{^{12}_{\Lambda}C}}$, and the second one from
proton kinetic energy spectra in ${\mathrm{^{5}_{\Lambda}He}}$,
${\mathrm{^{7}_{\Lambda}Li}}$, ${\mathrm{^{9}_{\Lambda}Be}}$,
${\mathrm{^{11}_{\Lambda}B}}$, ${\mathrm{^{12}_{\Lambda}C}}$,
${\mathrm{^{13}_{\Lambda}C}}$, ${\mathrm{^{15}_{\Lambda}N}}$ and
${\mathrm{^{16}_{\Lambda}O}}$. A branching ratio
for the $2N$-NM decay channel of this magnitude is consistent
with the prediction made by  Bauer and Garbarino~\cite{Bau09a}. On
the other hand,  the one-particle proton and neutron kinetic
energy spectra in ${\mathrm{^{4}_{\Lambda}He}}$ measured at
BNL~\cite{Pa07} are accounted for reasonably well theoretically by
considering only the $1N$-NM decay mode~\cite{Bau09,Kr14}.

The above mentioned experiments, together with several others performed during the last few
 decades~\cite{Sz91,No95,Ki03,Ok04,Ok05,Ou05,Ka06,Ki06,Bh07,Ag09},
represent very important  advances in  our knowledge about  NM decay.
 Explicitly, these advances are: 1) new
high quality measurements of  the number  of single-nucleons ${\rm N}_N$,
 as a function of the one-nucleon energy $E_N$,
and 2) first  measurements of the number of two-particle coincidences ${\rm N}_{nN}$,  as  a function of:
i) the sum of the kinetic energies $ E_{nN}\equiv E_n+E_N$,  ii) the opening angle
$\theta_{nN}$, and  iii) the center of mass (c.m.)  momentum
$P_{nN}=|\textbf{p}_n+\textbf{p}_N|$. On the theoretical side this  implies
a new challenge for nuclear models which
have to explain, not only the $1N$- and $2N$-NM decay rates, but also
the shapes and magnitudes of all these spectra, testing in this way both the
 kinematics and the dynamics.

Recently, Bauer, Garbarino,  Parre\~{n}o and  Ramos~\cite{Bau10,Bau10a,Bau11} have obtained good   agreement with   KEK data~\cite{Ki09},
considering both the one- and the  two-nucleon induced decays in the framework of the
 Fermi Gas Model (FGM). These authors have also analyzed the proton kinetic energy spectrum in
 ${\mathrm{^{12}_{\Lambda}C}}$
 measured at FINUDA ~\cite{Ag10}, but no theoretical analysis
 of the remaining spectra  has been done so far.
In the present work, we present for the first time the calculation of
all proton kinetic energy spectra  studied in the above mentioned
experiment.

Since i) the NMWD is dominated by the $1N$-NM decay, and ii)
the $2N$-NM processes and the FSIs contribute mainly at low energy,
it is reasonable and useful to compare the experimental spectra
with theoretical calculation when only the  $1N$-NM decay mode is considered.
\footnote{
 Such a comparison is analogous to those done between
the experimental data  on  electron-nucleus  and charged-current
neutrino-nucleus scatterings that
 include the FSIs, with the plane-wave impulse approximation
 which doesn't include these processes.~\cite{Be06,Me13}.
Moreover, the results of our analysis  are fully robust, in the sense
that they will be valid even after the inclusion of the FSIs  and
the $2N$-NM decay.}
It is obvious that there will be no agreement at low energies between the
experimental and theoretical spectra. But this disagreement 
 is not crucially important,
since we are basically interested in  disentangling  the strangeness-flipping interaction
among baryons~\cite{Kr14}.
 Of course, both the $2N$-NM decay  and  the FSIs, as well as the SRCs, are interesting 
 physical phenomena in themselves, but they teach us  little about the basic nonmesonic decay.
   Moreover, as indicated in \rf{3} they can not be treated separately, and it is not known whether they contribute coherently or incoherently. That is, it can even
   happen that they partially cancel out (for instance, $\Gamma_n^{0,FSI}$ and $\Gamma_{np}^{0,FSI}$
   terms in \rf{3}), as do the divergences and the vertex corrections in the QED, because of Ward identity.
 More specifically, and  as already pointed out in Ref.~\cite{Kr10a}, the issue
 of FSIs in the NMWD is a tough nut to crack, and there is no theoretical work in the
 literature encompassing all aspects of these processes.
 For this reason, before having a reasonable control over all the physics that they involve,
 it may be useful, or even preferable, to discuss the experimental data without the FSIs.
   This is what we do here.

The content of this article is as follows. Our method to compare the experimental data with theory for  the
NMWD is explained in detail in  Sec. \ref{Sec2}. The main formulas used to calculate the
 proton  spectra for the $1N$-NM decay within the Independent-Particle Shell Model (IPSM) is
presented in  Sec. \ref{Sec3}.
 The   parameterizations
that are used for vertices  are listed in Sec. \ref{Sec3} also. The only novelty here is that the
proposed  BBM vertex functions   are rarely used in the
literature. The calculated spectra for all hypernuclei are presented  in Sec. \ref{Sec4},
where we make a  comparison between theory and the FINUDA data for  ${\mathrm{^{5}_{\Lambda}He}}$,
${\mathrm{^{7}_{\Lambda}Li}}$, and  ${\mathrm{^{12}_{\Lambda}C}}$.
The  extraction  of $\Gamma^0_2$ from experimental spectra is reanalyzed in Sec. \ref{Sec5}.
 Finally, Sec. \ref{Sec6} presents the final remarks and conclusions.


\section{Relationship between Experiment and Theory}
\label{Sec2}

While $\Gamma_{N}$ and $\Gamma_{nN}$ are experimentally observable quantities,
 the bare decay rates $\Gamma^0_p$, $\Gamma^0_n$, and  $\Gamma^0_2$
are not and have to be derived from the data, employing different extraction procedures
which frequently involve approximations that are  questionable. (One example will be illustrated here.)
Moreover,  the primary decay rates  are ill defined, since they depend on the  model
that is used to describe the nuclear
structure of the  hypernucleus, as well as  on the SRC, etc.

Note that we include  the FSIs in the definition of $\Gamma_{N}$, and $\Gamma_{nN}$
which is not commonly done in the study of the NMWD~\cite{Bau10a,Kr10a}.
But,  there are other processes in nuclear physics
where the FSIs participate in the definition
 of the decay rates.  The best  known phenomenon is  nuclear
  $\beta$-decay, where the transition rate
  depends on the FSIs caused by the Coulomb attraction of the emitted electrons.
 (See, for instance, Eq. (5.11) in Ref.  \cite{Sc66} where  the FSIs effects
 are approximated by the Fermi function.)
  The main difference between the FSIs in the leptonic and nonmesonic weak-decays is that
 while in the first case  they are easily evaluated,
 in the second case  they are very complicated~\cite{Kr10a} and beyond the scope of this work.

The FSIs are usually  simulated by a semi-classical
 model,  developed by Ramos \etal~\cite{Ra97}, and denominated   Intranuclear Cascade (INC) code.
 This code interrelates the primary rates \rf{2} with measured rates \rf{3}.
More recently, the FSIs were evaluated with  a time-dependent
multicollisional Monte Carlo cascade scheme~\cite{Go11,Go11a}, implemented within the CRISP
code (Collaboration Rio-S˜ao Paulo), which describes, in a phenomenological
way, both the nucleon-nucleon scattering inside the nucleus and the escape
of nucleons from the nuclear surface
\cite{CRISP1,CRISP2,CRISP3,CRISP4}.
The CRISP code, as all INC codes, is  tailored
to simulate the experimental data, and not for describing theoretically
 the FSIs. As  such, it involves  a normalization procedure
for the experimental data, which  washes out all of the information
on the NMWD dynamics. More specifically, the  spectra of
$\Gamma_1^0$ evaluated from a Shell Model (SM)
 are the main ingredients for establishing the initial condition to start the CRISP cascade
process and to calculate in this way the  FSIs. (The primary
spectra of $\Gamma_2^0$  should  also be included within the
initial conditions in a more complete model, but we do not know
yet how are they evaluated within the SM.)
Yet, because of the normalization,  the spectrum perturbed
 by the FSIs  turns out to be  the same for different primary spectra. (One obtains the
 same $\Gamma_p^{0,FSI}$  spectra for different $\Gamma_p^{0}$ spectra, etc.)
We were not  able, so far, to find out how to circumvent this problem of normalization.
Moreover, not all FSIs are considered within the INC codes. Which additional FSIs
contribute to the NMWD spectra and decay rates, and
how and which of them should be included in the calculation are
 nontrivial questions.  Some candidates are discussed in Ref. \cite{Kr10a}.

The information on the dynamics also is lost  when the
 decay rates $\Gamma_{N}$, and $\Gamma_{nN}$ are normalized to  $\Gamma_{NM}$.
 This normalization is the usual procedure;
 see for instance
~\cite[Eq. (7)]{Bau10}, and ~\cite[Eqs. (14),(15)]{Ga04}.
With this normalization, the spectra depend on the phase space and the FSIs,
but very weakly on the  NMWD dynamics.
 The same happens when the transition density is normalized to the decay rate
 (See ~\cite[Fig. 3]{Ba08}.) 
In contrast, as explained below,  we do not  normalize any of the calculated transition rates,
and this  allows us to inquire more deeply into the NMWD mechanism.
\footnote{It might be useful to draw a parallel with electromagnetic decay.
When the electric $E2$ and the magnetic $M1$ multipoles are the lowest allowed transitions,
both may contribute significantly to the total rate $\Gamma_\gamma=\Gamma_\gamma(E2)+\Gamma_\gamma(M1)$,
since the electric transition may be enhanced
substantially above the single particle estimate due to collective effects.
The comparison between theory and data is always done for the   decay rates
$\Gamma_\gamma(E2)$ and
$\Gamma_\gamma(M1)$, separately. There is no physical motivation for comparing
the ratios $\Gamma_\gamma(E2)/\Gamma_\gamma$ and $\Gamma_\gamma(M1)/\Gamma_\gamma$,
 since they are less sensitive to the nuclear structure effects than the individual decay rates.
 }

The measurement implies the  counting of the numbers of emitted protons $\Delta
{\rm N}^i_p$ and the errors $\delta\Delta {\rm N}^i_p$,
 corrected by the detection efficiency,  within $m$  energy bins of width  $\Delta E$.
The total number of emitted protons and the resulting errors are
  \ber
{\rm N}_{p}=\sum_{i=1}^m\Delta{\rm N}^i_p,
\hspace{1cm}\delta{\rm N}_{p}=\sqrt{\sum_{i=1}^m(\delta\Delta{\rm N}_{p}^i)^2},
 \label{5} \eer
where  the summation goes over proton energies $E^i_p$ larger than a given
threshold energy $E_{thres}$.

The corresponding total decay rate with its error read
 \ber
\Gamma_{p}=\sum_{i=1}^m\Delta\Gamma_p^i,
\hspace{1cm}\delta\Gamma_{p}
=\sqrt{\sum_{i=1}^m(\delta\Delta\Gamma_p^i)^2},
 \label{6} \eer
where, as seen from \rf{4}, the  decay rates $\Delta \Gamma^i_p(E_p)$ with errors
$\delta\Delta\Gamma_p^i$ are given by
\ber
\Delta\Gamma_p^i(E_p)=\frac{\Gamma_W}{{\rm N}_W}
\Delta{\rm N}^i_p(E_p),
  \label{7}\eer
 and
\br
\delta\Delta\Gamma_p^i
&=&\frac{\Gamma_W}{{\rm N}_W}
\Delta {\rm N}^i\left[ \left(\frac{\delta\Delta {\rm N}^i}{\Delta {\rm N}^i}\right)^2
 +\left(\frac{\delta\Gamma_W}{\Gamma_W}\right)^2+\left(\frac{ \delta N_W}{N_W}\right)^2\right]^{1/2},
\label{8}\er
with  $\delta\Gamma_W$, and  $\delta N_W$
being, respectively,  the experimental errors on  $\Gamma_W$, and  $N_W$.
Thus,  to evaluate the experimental decay rates
 we need to know the values of $\Gamma_W$ and ${{\rm N}_W}$ for each hypernucleus.
 As  usually done all $\Gamma$'s will be
given in units of $\Gamma_\Lambda$, the total decay width of the free $\Lambda$.

For the first quantity we can  use the relationship
\be
{\Gamma}_{W}(A)=(0.990\pm 0.094)+(0.018\pm0.010)~A,
\label{9}\ee
which was derived in Ref.~\cite{Ag09} from a linear fit to the known values  of
 all measured hypernuclei in the mass range $A=4-12$.


But unfortunately there is no experimental information about
${\rm N}_W$ in the literature.
 Only the ratio
 \be
  R_p=\frac{{\rm N}_{p}}{{\rm N}_{W}},
\label{10}\ee
for the eight hypernuclei discussed here were presented at a conference ~\cite{Bot14},
 for the threshold energy $E_{thres}=15$ MeV.
  We will use, however, only the results for
${\mathrm{^{5}_{\Lambda}He}}$, ${\mathrm{^{7}_{\Lambda}Li}}$, and
${\mathrm{^{12}_{\Lambda}C}}$,   since only these were published in
a  refereed physics journal so far~\cite{Ag08}.
We list them  in Table \ref{T1},
  together with values of ${\rm N}_{p}$  from Ref.~\cite{Ag10}, and the  resulting estimates for  ${\rm N}_W$
  from \rf{10}.  The proton decay rates, evaluated  from
    \ber
\Gamma_{p}=R_p{\Gamma_{W}},
  \label{11}\eer
are also shown.
Since, according to \rf{9}, the value of ${\Gamma}_{W}$ is close to unity, the last result implies
that the ratio $R_p$ is basically the proton decay rate.

\begin{table}[htpb]
\caption{
Values of $R_{p}$  from Ref.~\cite{Ag08} and ${\rm N}_{p}$  from Ref.~\cite{Ag10} for $E_{thres}=15$ MeV, together with the resulting estimates for  ${\rm N}_W$, and  $\Gamma_p=R_p\Gamma_{W}$.}
 \label{T1}
\bigskip
\begin{center}
\begin{tabular}{ccccc}
\hline\noalign{\smallskip}
Hypernucleus    &$R_{p}$&${\rm N}_{p}$
&${\rm N}_{W}$& $\Gamma_{p}$\\
\noalign{\smallskip}
\hline\noalign{\smallskip}
$_\Lambda^{5}$He &$0.25\pm  0.07 $&   $262\pm    25$   & $      1047\pm       391$&$ 0.27\pm   0.11$\\
$_\Lambda^{7}$Li &$0.37\pm  0.09 $&   $259\pm    21$   & $       700\pm       226$&$  0.41 \pm  0.16$\\
$_\Lambda^{12}$C &$0.43\pm  0.07 $&   $678\pm    38$   & $   1576\pm   344$&$   0.52 \pm  0.18$\\
\noalign{\smallskip}
\hline
\end{tabular}
\end{center}
\end{table}
The theoretical analogs  of \rf{6} and \rf{7} are, respectively
  \be
\Gamma^{th}_{p}=\int S_p(E_{p})dE_p,
\label{12}\ee
and
  \be
\Delta \Gamma^{th}_p(E_p)=S_p(E_p)\Delta{E},
\label{13}\ee
where   the spectral function $S_p(E_p)$ depends on the theory that  is used to evaluate
the NMWD, which not yet has been discussed.

 Instead of comparing  the experimental transition rates with the calculated rates,
 we can compare directly  the  number of measured protons  with the calculated quantity
\be
\Delta {\rm N}^{th}_p(E_p)=\frac{{\rm N}_W}{\Gamma_W}S(E_p)\Delta{E},
\label{14}\ee
where ${\rm N}_W/{\Gamma_W}$ is just a proportionality factor.

All the above is completely general.
The theory that is used can be as complicated as necessary to properly interpret experimental
 data. But it can also be very simple and still  lead us to correct conclusions
 about the underlying physics.
\footnote{The  model used to describe a process should, in principle, describe all the involved physics.
 This is a desirable condition, but it is not in any way essential.
Useful models are those that allow us to infer consequences consistent with the observations. More precisely, a model is a simplified version of the process, and the model designer  decides which features to consider.}
Such a model is described below.


\section{ Independent Particle Shell Model for  the Spectral Function}
\label{Sec3}

The IPSM  has been  used for more than twenty years in the evaluation of the
$1N$-NM decay rates~\cite{He86,Co95}, but only in recent years
 was it applied for the description of  different spectral densities $S_{N}(E)$,
  $S_{nN}(E)$, $S_{nN}(\cos\theta)$, and  $S_{nN}(P)$ ~\cite{Bau09,Kr14,Kr10a,Go11,Go11a,Ba02,Kr03,Ba03,Ba07,Ba08,Ba10,Kr10}.
We briefly sketch here the main assumptions that are made in this model, and
give the resulting theoretical expression for the proton
kinetic energy spectrum.

The assumptions are: (i)
 the initial hypernuclear state is taken as a hyperon $\Lambda$ in a
single-particle state $j_\Lambda=1s_{1/2}$ weakly coupled to an
$(A-1)$ nuclear core of spin $J_C$, i.e.,
$\ket{J_I}\equiv\ket{(J_Cj_\Lambda)J_I}$; (ii)
the nucleon ($N=p,n$) inducing the decay  is in the single-particle state
$j_N$ ($j\equiv nlj$); (iii) the final  residual nuclear states are:
$\ket{J_F}\equiv\ket{(J_Cj_N^{-1})J_F}$; (iv)
 the liberated energy is
\be \Delta^j_{N} =\Delta  + \varepsilon_{\Lambda} +
\varepsilon^j_{N},
 \label{15}\ee
where  $\Delta=\Mass_\Lambda-\Mass_p=177.33$ MeV, and the $\varepsilon$'s are
experimental single-particle energies (s.p.e.)
\footnote{The schematically drawn energies in~\cite[Fig. 7]{Kr10} are  the experimental s.p.e.,
which  can be identified with the SM s.p.e. only  in closed shell nuclei, such as ${\mathrm{^{16}O}}$.
For open shell nuclei, which is the case of ${\mathrm{^{12}C}}$, the
experimental s.p.e. are frequently  identified with the quasiparticle energies, which include
the effect of pairing correlations, and could be quite different from the SM s.p.e..
 This was done, for instance,  in ~\cite[Table IV]{Krm05}, where the experimental and SM s.p.e.
  are listed, respectively, in columns two and five.
  Moreover, the correct $p_{3/2}$,  $p_{1/2}$ and $d_{5/2}$ experimental energies
  read, respectively,  $-15.96$,  $-1.95$, and $1.61$ MeV
~\cite{TUNL}.}
, and (v) the c.m. momenta and relative momenta  of the emitted particles
are:
 \ber
P_{nN}&=&\sqrt{(A-2)(2\Mass\Delta^j_{N}- p_n^2 -p_N^2)},\nn\\
~p_{nN}&=&\sqrt{\Mass \Delta^j_{N}- \frac{A}{4(A-2)} P_{nN}^2}.
\nn
\eer
It follows that the $1N$-NM decay rate is given by~\cite{Kr10}
 \[
\Gamma_{N}=\sum_{j}\Gamma_{N}^j;~~~\Gamma_{N}^j=\int
\I^j_{N}(p_{nN},P_{nN})d\Omega_{nN},
\]
 where $d\Omega_{nN}$ is the phase space factor,
 and
 \brn
&&\I^j_{N}(p,P)=\frac{1+\delta_{Nn}}{2}\sum_{J=|j-1/2|}^{J=j+1/2}
F^{\sf j}_{NJ}
\sum_{SlL\lambda T}|\M(plPL\lambda SJT;{j_\Lambda {\sf j}_NJT})|^2,
\ern
with
\brn
\M(plPL\lambda SJT;j_\Lambda j_N JT)
&=&\sqi\left[1-(-)^{l+S+T}\right]
\nonumber \\
&\times&\O_L(P)
({lL\lambda SJT}|{ V}(p)|{j_\Lambda j_N JT}),
\ern
and
 \[
 \O_L(P) =\int R^2dRj_L(PR){\rm R}_{0L}(b/\sqrt{2},R).
 \]
Here $L$,  $l$, and $\lambda$ are, respectively, the c.m., relative, and total orbital angular
momenta ($\mblambda=\mbL+\mbl$), while $V$ is the transition potential, and  $b$
is the harmonic oscillator length parameter.

The kinematics of different spectra $S_N$
depend on  $d\Omega_{nN}$ and on the overlap $\O_L(P_{nN})$,  while
the decay dynamics is contained in the matrix element $({lL\lambda SJT}|{ V}(p_{nN})|{j_\Lambda j_N JT})$.
The information on  nuclear structure is
enclosed in the spectroscopic factors $F^j_{NJ}$, which account for the Pauli Principle within
each single-particle shell $j_N$. In the general case, they are given by
\br F^j_{NJ}&=&(2J+1)
\sum_{J^n_F}\sixj{J_C}{J_I}{j_\Lambda}{J}{j_N}{J_F}^2
|\Bra{J_C}a_{j_N}^\dag\Ket{J^n_F}|^2,
 \label{16}\er
where
$\Bra{J_C}a_{j_N}^\dag\Ket{J^n_F}$ are the fractional parentage coefficients (FPCs), and
 the summation goes over the  $n$ final states $J^n_F$
in the residual $(A-2)$ nuclei,  with the same spin, parity, and isospin, and different excitation energies.
Such a detailed description could  be redundant since, in the NMWD,  we evaluate the  inclusive decay
rate, without being interested in exclusive processes that feed each of the individual final states $J^n_F$.
Thus, within the IPSM, the spectroscopic factor becomes much  simpler since $\ket{J^n_F}\go\ket{(J_Cj_N^{-1})J_F}$,
 and the summation  goes only over the values of $J_F$ that fulfill the constraint $|J_C-j_N|\le J_F\le J_C+j_N$.
 The values for $J_I$ and $J_C$ are  taken from experimental data and, for the hypernuclei of interest here, are
listed in Table I of Ref. \cite{Kr10}. The resulting factors $F^j_{NJ}$ are listed in Table II of the same paper.

 The  spectra are obtained from the differentiation of  $\Gamma_N$
 with respect to $E_N$, $\cos\theta_{nN}$,  $E_{nN}$, and  $P_{nN}$. In particular,
 for the kinetic energy spectrum, one has:
\be
S_N(E_{N})=\sum_{j}S_N^j(E_{N}),
\label{17}\ee
with
\brn
S_N^j(E_{N})&=&(A-2)\frac{8\Mass^3}{\pi}\int
_{-1}^{+1} d\cos\theta_{nN}
\sqrt{\frac{E_{N}}{E_N'}}\, E_n\,
\I_{j_N}(pP),\ern
where
 \ber
E'_N&=&(A-2)(A-1)\Delta^j_{N}-E_{N} [(A-1)^2-\cos^2\theta_{nN}],
\nn\eer
and
  \ber
  E_n&=&\left[\sqrt{E'_N} -\sqrt{E_{N}}\cos{\theta_{nN}}\right]^2(A-1)^{-2}.
\nn\eer
Finally,
 \ber
\Gamma_{N}\equiv\sum_{j}\Gamma_{N}^j=\sum_{j}\int_0^{{Q}^j_{N}}S_N^j(E_{N})dE_N,~~~~{Q}^j_{N}=\frac{A-2}{A-1}\Delta^j_{N},
 \label{18}\eer
with ${Q}^j_{N}$ being the single-particle Q-values.

 The  outline of the numerical calculation is the following:

\begin{enumerate}

\item  The transition potential
  $V(p_{nN})$ for  the emission of the  $nN$ pair,  contained in $\T_{NJL}^j(p_{nN})$,
is  described  by three  OME models, namely: P1 - The full pseudoscalar ($\pi, K, \eta$) and vector
($\rho,\omega,K^*$) meson octets (PSVE),  with the weak coupling
constants, and dipole form-factor cutoffs $\Lambda_M$ from Refs.~\cite{Du96,Pa97,Pa02};
P2 - Only one-$(\pi+ K)$ exchanges (PKE) are considered, with the same parametrization as
in the previous case, \ie  with cutoffs $\Lambda_\pi= 1.3$ GeV and  $\Lambda_K=1.2$ GeV
from~\cite{Pa97}; and P3 - The soft $\pi+ K$ exchange (SPKE) potential with cutoffs $\Lambda_\pi= 0.7$ GeV and
$\Lambda_K=0.9$ GeV from~\cite{Bau09,Kr14}.

\item The SRCs acting on
 final  $nN$ states are incorporated
 phenomenologically  through Jastrow-like SRC functions, as used
  within both finite nuclei calculations
 ~\cite{Pa97,Ba02,Kr03,Ba03,Ba07}, and Fermi Gas Model (FGM) calculations
 ~\cite{Bau09a,Bau09b}.

 \end{enumerate}


\section{Proton Decay Rates and  FINUDA data}
\label{Sec4}

\begin{figure}[th]
\begin{center}
\resizebox{1.1\textwidth}{!}{
\begin{tabular}{cc}
\psfig{file=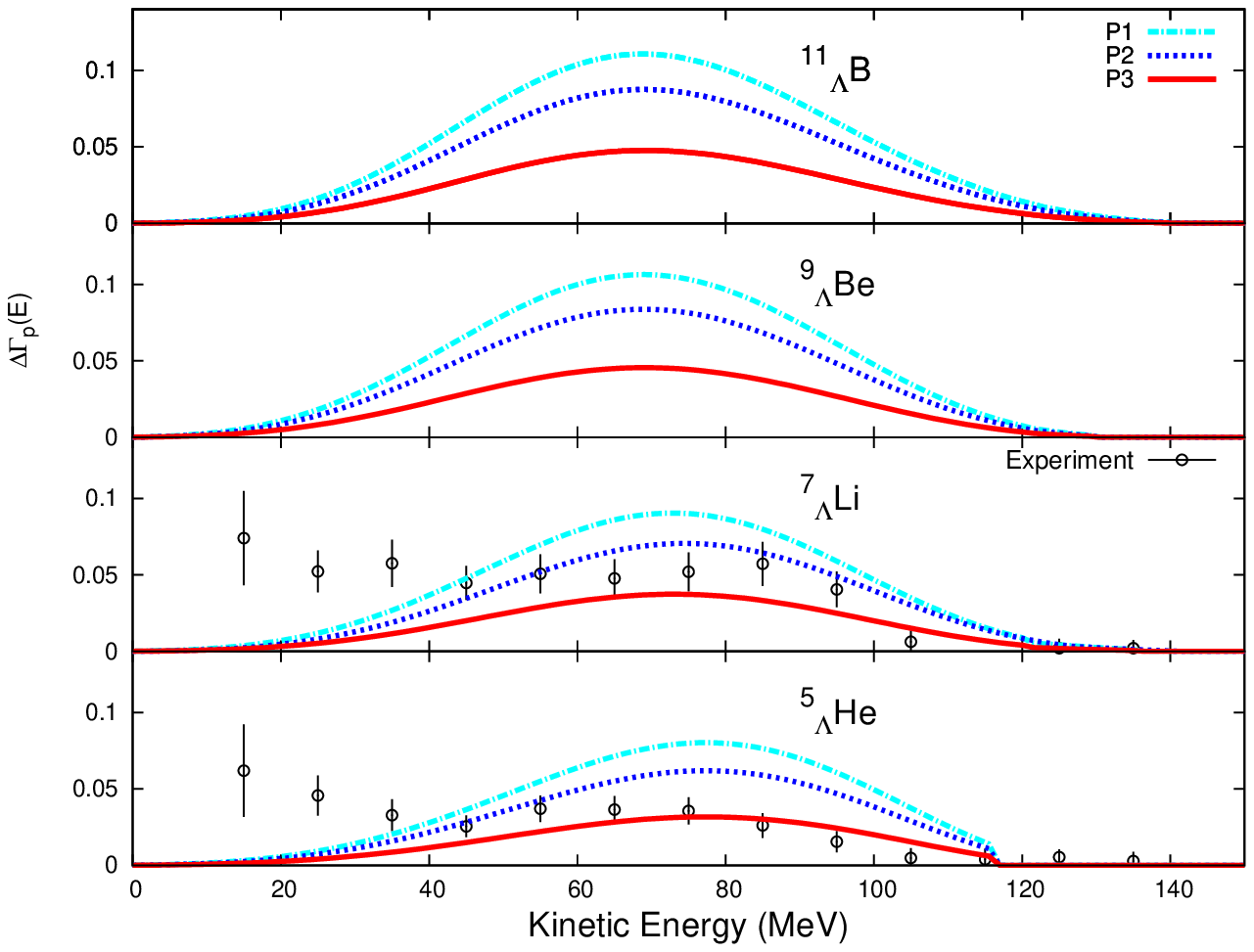,width=8cm,height=10cm} &
\psfig{file=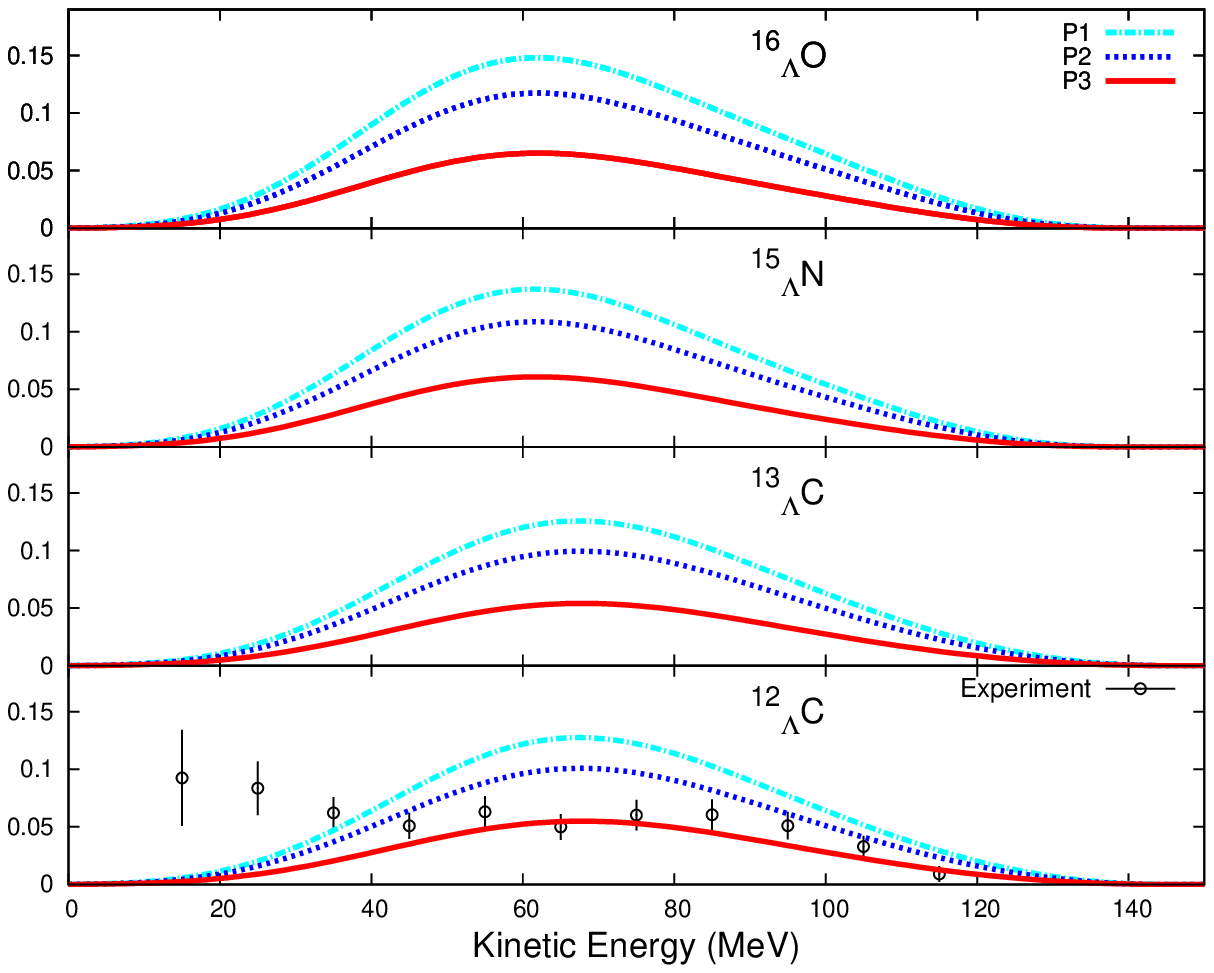,width=8cm,height=10cm}
\end{tabular}}
\vspace{-0.5cm}
\caption{\label{Fig2}(Color online) Calculated
proton kinetic energy spectra  $\Delta \Gamma^{th}_p(E_p)$
for three different transition potentials specified
 in the previous section. For ${\mathrm{^{5}_{\Lambda}He}}$,
${\mathrm{^{7}_{\Lambda}Li}}$, and ${\mathrm{^{12}_{\Lambda}C}}$ are also shown
 the values of $\Delta \Gamma_p(E_p)$ evaluated from \rf{7} that correspond to the
$\Delta{\rm N}_p(E_p)$ shown in~\cite[Fig. 1]{Ag10}.}
\end{center}
\end{figure}

 The calculated transition densities  $\Delta \Gamma^{th}_p(E_p)$, for  $\Delta E=10$ MeV and
evaluated from  \rf{13} with  $S_p(E_p)$ given by \rf{17}
are shown in Fig. \ref{Fig2} for the hypernuclei measured by
FINUDA\cite{Ag10}.
As expected,   in all cases the spectra strongly depend on the parameterization that is
used for  the transition potential, while  their shapes are the same for all practical purposes.
The experimental values of $\Delta \Gamma_p(E_p)$ for ${\mathrm{^{5}_{\Lambda}He}}$,
${\mathrm{^{7}_{\Lambda}Li}}$, and ${\mathrm{^{12}_{\Lambda}C}}$, evaluated from
\rf{7} for the values of $\Delta{\rm N}_p(E_p)$ shown in~\cite[Fig. 1]{Ag10},
are also displayed in the same
figure. Their errors were calculated from \rf{8}.

With the parametrization P1, the theory greatly overestimates
the experimental spectra  at medium and high energies ($E_p\gsim 40$ MeV), underestimating them
at low energies.
The discrepancy can not be settled by simply opening a new $2N$-NM
decay channel induced by two nucleons, since this decay mode,
although capable of producing additional particles at low
energies, is unable to lower the transition strength at high energies.
Nor is it possible for the FSIs to solve the problem, since they can hardly change the total  transition density
 induced by a proton. They can only  remove  a portion of the strength  from
 high energy to low energy. It is self-evident from Fig. \ref{Fig2} that such a mechanism
 cannot be successful in the present case.

Improved  agreement is obtained
in the P2 model, which means that
 the incorporation of vector mesons, instead of improving the agreement,
 makes it poorer. However, the high energy part of the ${\mathrm{^{5}_{\Lambda}He}}$
 spectrum is reproduced fairly well
only with the parametrization P3. The decrease in magnitude of proton spectra in going
from P1 to P2 is due to the well known fact that the parity-violating
contribution of the vector mesons to the  proton transition rates is quite
sizable (see, for instance,~\cite[Table IV]{Ba02}). On the other hand,
the strong variation of  the same observable  in $_\Lambda^{5}$He  with regard
to the BBM vertex functions is thoroughly discussed in ~\cite[Fig. 1]{Bau09}.

\begin{table}[htpb]
\caption{Transition rates $\Gamma_{p}^{th}$ calculated  from \rf{12}
 for all three parametrizations and with two different threshold energies:
 $E_{thres}=15$ MeV,  and $E_{thres}=40$ MeV.
 For ${\mathrm{^{5}_{\Lambda}He}}$,
${\mathrm{^{7}_{\Lambda}Li}}$, and ${\mathrm{^{12}_{\Lambda}C}}$
 the values of $\Gamma_p$ and their errors,  evaluated from \rf{6}, are also shown.
} \label{T2}
\begin{center}
\begin{tabular}{ccccc}
\hline\noalign{\smallskip}
Hypernucleus &$\Gamma_{p}$&$\Gamma_{p}^{\mbox{\tiny P1}}$&
$\Gamma_{p}^{\mbox{\tiny P2}}$&$\Gamma_{p}^{\mbox{\tiny P3}}$
\\
\noalign{\smallskip}
\hline
\noalign{\smallskip}
\underline{$E_{thres}=15$ MeV}&&&&\\
\noalign{\smallskip}

$_\Lambda^{5}$He &$0.27\pm 0.05 $&$0.466$&$0.360$&$0.185$\\
$_\Lambda^{7}$Li &$0.41\pm 0.07 $&$0.531$&$0.415$&$0.228$\\
$_\Lambda^{9}$Be &&$0.627$&$0.494$&$0.275$\\
$_\Lambda^{11}$B &&$0.667$&$0.527$&$0.289$\\
$_\Lambda^{12}$C &$0.52\pm 0.07 $&$0.792$&$0.627$&$0.343$\\
$_\Lambda^{13}$C &&$0.776$&$0.614$&$0.336$\\
$_\Lambda^{15}$N &&$0.821$&$0.651$&$0.365$\\
$_\Lambda^{16}$O &&$0.906$&$0.718$&$0.399$\\
\noalign{\smallskip}
\hline
\noalign{\smallskip}
\underline{$E_{thres}=40$ MeV}&&&&\\
\noalign{\smallskip}
$_\Lambda^{5}$He &$0.19\pm 0.04 $&$   0.428$&$   0.331$&$   0.171$\\
$_\Lambda^{7}$Li &$0.27\pm 0.05 $&$  0.488 $&$  0.381 $&$  0.204$\\
$_\Lambda^{9}$Be &&$   0.555 $&$  0.437$&$   0.243$\\
$_\Lambda^{11}$B &&$   0.607 $&$  0.481$&$   0.263$\\
$_\Lambda^{12}$C &$0.37\pm 0.06 $&$  0.719 $&$  0.569 $&$  0.311$\\
$_\Lambda^{13}$C &&$  0.706 $&$  0.559 $&$  0.305$\\
$_\Lambda^{15}$N &&$  0.721 $&$  0.573 $&$  0.320$\\
$_\Lambda^{16}$O &&$   0.801 $&$  0.635 $&$  0.352$\\
\noalign{\smallskip}
\hline
 \end{tabular}
\end{center}
\end{table}

At this stage, it might be useful to recall that, in all theoretical descriptions of the NMWD,
only three body final states have been considered, implying that the residual
nuclei are necessarily bound. Obviously, this is done for simplicity, but
 does not always occur. The most emblematic case is that of ${\mathrm{^{11}_{\Lambda}B}}$,
  which has been considered in many theoretical studies done so far.
  However, its parent nucleus in the neutron channel, ${\mathrm{^{11}B}}$,
  is unstable and disintegrates into
 $p+2\alpha$ with a half-life of $8\cdot 10^{-19} s$, which is very  short when compared
 with the half-life of ${\mathrm{^{11}_{\Lambda}B}}$. Among the nonmesonic decays
  analysed here,  the same occurs with the ${\mathrm{^{5}He}}$ nucleus,
which is the residual nucleus for  the proton NMWD of ${\mathrm{^{7}_{\Lambda}Li}}$.
In fact, it is unstable
to particle emission, decaying into $p + {\mathrm{^{4}He}}$ with a half life of
$70(3)\cdot 10^{-25} s$.
 This time is much shorter than  the lifetime of ${\mathrm{^{7}_{\Lambda}Li}}$ and,
therefore, the instability  of ${\mathrm{^{5}He}}$ could be the cause of the
discrepancy between theory and experiment at high energy.
\begin{table}[htpb]
\caption{Spectroscopic factors and the transition rates $\Gamma_{p}$
for ${\mathrm{^{7}_{\Lambda}Li}}$ evaluated in the intermediate coupling model with
the wave function amplitudes $a_{LS}$  given in Ref.~\cite{Mil07}, which are also listed.} \label{T3}
\begin{center}
\begin{tabular}{crrrrr}
\hline\noalign{\smallskip}
\underline{$a_{LS}$ }&&&&&\\
\noalign{\smallskip}
$LS$  &$fit69$&$fit5$&$CK616$&$CK616$&CKPOT\\
\noalign{\smallskip}
\hline
\noalign{\smallskip}
$01        $&$0.9873 $&$0.9906$&$ 0.9576$&$ 0.9484$&$ 0.9847$\\
$12        $&$-0.0422$&$ -0.0437$&$ -0.2777 $&$-0.3093$&$ -0.1600$\\
$10        $&$ -0.1532$&$ -0.1298$&$ -0.0761$&$ -0.0703$&$ -0.0685$\\
\noalign{\smallskip}
\hline

\noalign{\smallskip}
\underline{$F^{j}_{NJ}$}&&&&&\\
\noalign{\smallskip}
$J,lj$ &&&&&\\
\hline
\noalign{\smallskip}
$ 0 , p_{1/2}  $&$       0.0320  $&$    0.0311 $&$     0.0693  $&$    0.0790  $&$    0.0419$\\
$ 1,  p_{1/2}  $&$       0.3023  $&$    0.3032 $&$     0.3026   $&$   0.3022 $&$     0.3042$\\
$ 1,   p_{3/2}  $&$         0.5255  $&$    0.5259  $&$    0.4973   $&$   0.4903  $&$    0.5175$\\
$ 2 ,  p_{3/2}   $&$       0.1403 $&$     0.1399  $&$    0.1307$&$      0.1286  $&$    0.1364$\\
\hline
 $\Gamma_p$&$ 0.2444 $&$      0.2444$&$      0.2475$&$                    0.2483  $&$0.2453$\\
\noalign{\smallskip}
\hline
 \end{tabular}
\end{center}
\end{table}

Similarly to what was done in Figure \ref{2} for proton kinetic energy spectra  $\Delta \Gamma^{th}_p(E_p)$,
the theoretical results for the proton decay rates (18)
 for two different threshold energies
 ($E_{thres}=15$ MeV,  and $E_{thres}=40$ MeV)  are shown  in Table \ref{2}.
 For ${\mathrm{^{5}_{\Lambda}He}}$,
${\mathrm{^{7}_{\Lambda}Li}}$, and ${\mathrm{^{12}_{\Lambda}C}}$
 the experimental values of $\Gamma_p$  and their errors, evaluated
 from \rf{6}, are also shown.

All that was previously stated when comparing the theory with the data in Figure \ref{2} also applies here.
 In particular, since the theory does not include the 2N-NM channel, the experimental transition rates must
always be larger than the theoretical
rates for  $E_{thres}=15$ MeV.
This condition is satisfied only for the parameterization P3, suggesting that
the other two sets of parameters would not be appropriate.
On the other hand, as the FSIs, which remove the density transition from the  high energy region,
are omitted in the calculations, all three $\Gamma^{th}_{p}$ should be
 larger than $\Gamma_{p}$ for $E_{thres}=40$ MeV. From Table \ref{T2}, we see that within the experimental errors this is indeed the case.
However, in calculations P2 and P3, the differences between the data and the theory are too large to
be entirely attributed to the lack of the FSIs in the theory.

It would also be interesting to compare the  experimental results shown in the upper  part
 of Table \ref{T2} with  the
calculation done by Itonaga and Motoba~\cite{It10} with a more elaborate SM than that used here,  employing
 the $\pi+2\pi/\rho+2\pi/\sigma+\omega+K+\rho\pi/a_1+\sigma\pi/a_1$ exchange potential. They obtain:
$\Gamma^{0}_{p}(_\Lambda^{5}$He)$=0.237$, and $\Gamma^{0}_{p}(_\Lambda^{12}$C)$=0.534$.
In the first case, $\Gamma_{p}$ is greater than $\Gamma^{0}_{p}$ as it should be, but it is lower in the second case which is not correct.

\begin{figure}[th]
\begin{center}
\resizebox{1.1\textwidth}{!}{
\begin{tabular}{cc}
\psfig{file=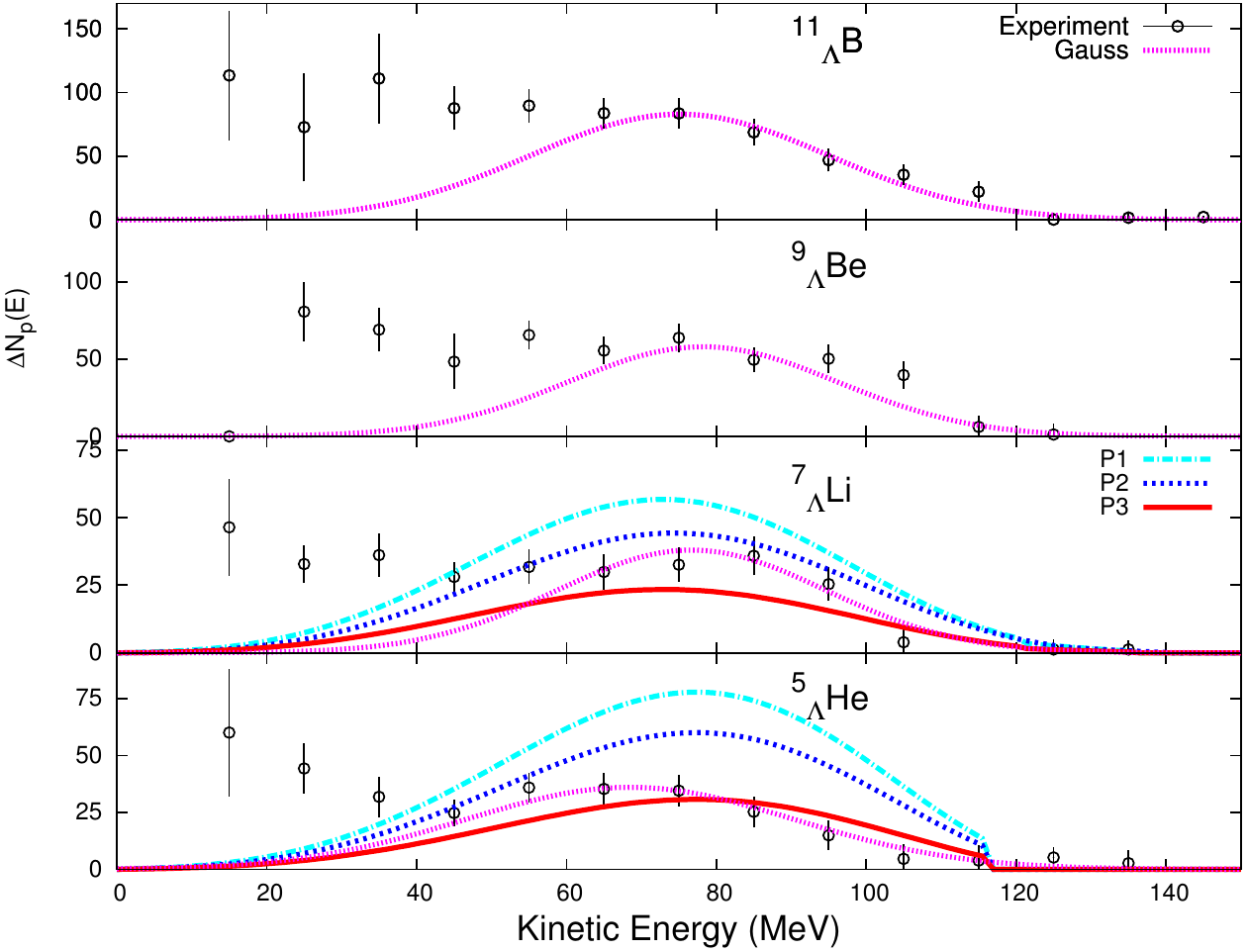,width=8cm,height=10cm} &
\psfig{file=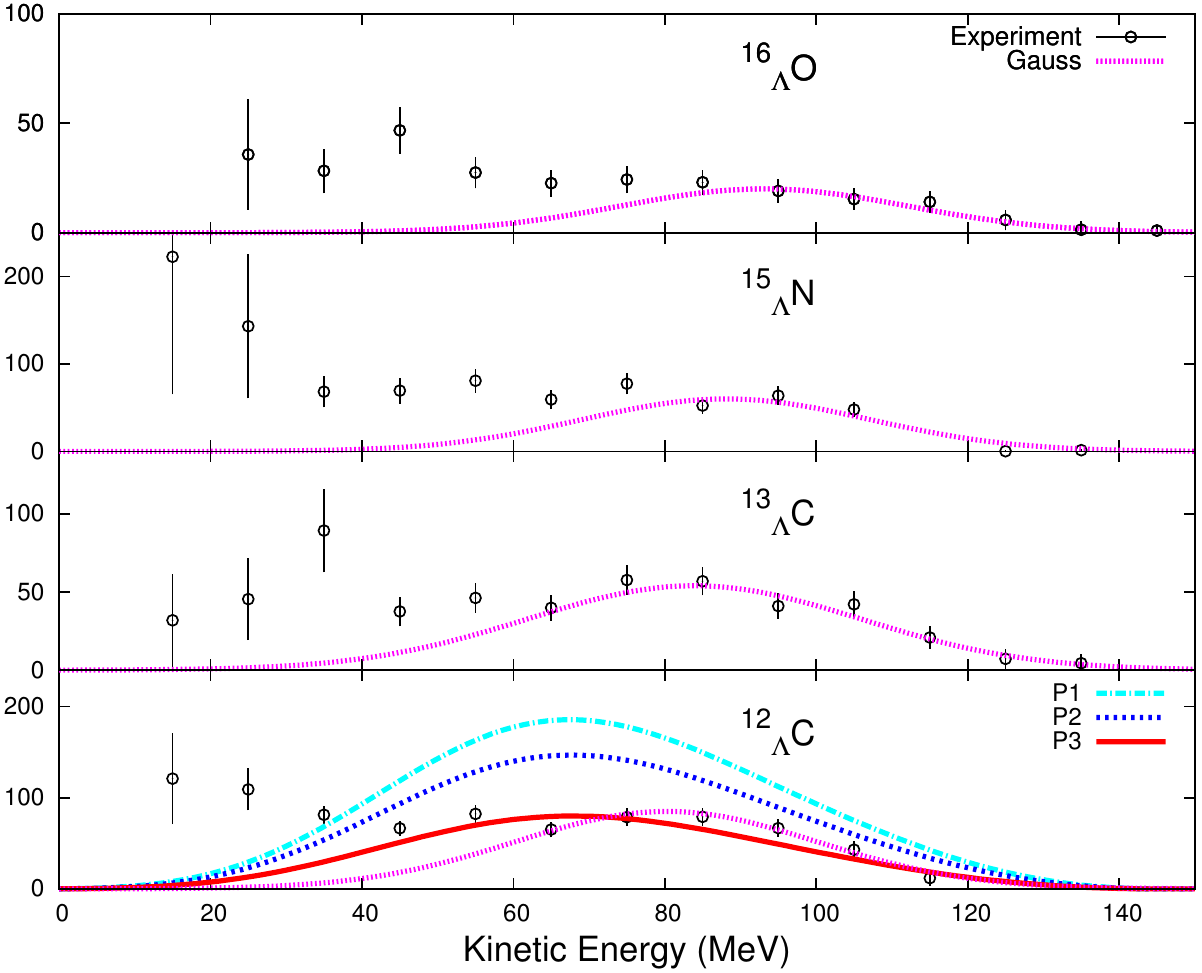,width=8cm,height=10cm}
\end{tabular}}
\caption{\label{Fig3}(Color online) Experimental
data~\cite{Ag10} for proton kinetic energy spectra for the number
of particles $\Delta {\rm N}_p(E_p)$ are compared with the IPSM
results for three different OME potentials.
 Theoretical results have been evaluated from equation \rf{14}.
 Also  shown are the Gaussian-function fits
of  each proton spectrum from $80$ MeV onwards, performed in
Ref.~\cite{Ag10}.}
\end{center}
\end{figure}

It can  be argued  that a derivation of the $F^j_{NJ}$, based on
the $jj$ single-particle model,
is not fully appropriate for light nuclei, such as Li and
Be, with the intermediate coupling model  preferred over  pure $jj$
coupling for the core nuclei.
In fact, the structure of these nuclei is closer to $LS$ coupling than to
an assembly of $p_{3/2}$  valence nucleons, as can be seen, for instance, from~\cite[Table 5]{Mil07}
where the core state $J_C=1^+$ in  ${\mathrm{^{7}_{\Lambda}Li}}$ is basically a pure $^3S$ state.
In this case, instead of the spectroscopic factors $F^{p_{3/2}}_{N1}=5/8$, and $F^{p_{3/2}}_{N2}=5/8$ given in Ref.~\cite[Table 2]{Kr10},
one has $F^{p_{3/2}}_{N1}=19/36$, $F^{p_{3/2}}_{N2}=5/36$, $F^{p_{1/2}}_{N0}=1/36$, and $F^{p_{1/2}}_{N1}=11/36$.
Within the parametrization P3, the last spectroscopic factors yield
$\Gamma_p=0.244$ which  is only  slightly larger that the value $0.228$ shown in Table~\ref{T1}.
However, to justify even more reliably the $jj$ coupling,
 we have
recalculated  $\Gamma_p$ with five different wave functions  evaluated in the intermediate-coupling model, and
cited in \cite{Mil07}.
 Their amplitudes $a_{LS}$ are listed
 in Table \ref{T3}. The spectroscopic factors \rf{6} are evaluated from the expression
\br
 F^{j}_{pJ}&=&3(2J+1)(2j_p+1)\sum_{J_F=1/2,3/2}(2J_F+1)\sixj{1}{\fot}{\fot}{J}{j_p}{J_F}^2
\nn\\
&\times& \sum_{LS}(2L+1)(2S+1)\left[a_{LS} \ninj{1}{\fot}{j_p}{1}{\fot}{J_F}{L}{S}{1}\right]^2.
\label{19}\end{eqnarray}
The results are shown in Table \ref{T3}, from which it can be concluded
 that the difference between the $jj$ coupling and the intermediate coupling
 is  $\leq 10\%$. The physical reason for this fact is that $\Gamma_p$ is an
inclusive quantity, so it is not acutely relevant if the proton decays from orbital $p_{3/2}$ or $p_{1/2}$.

We would like to stress that the way to compare the theory with  data
 as done here, as well as in a previous paper
  ~\cite{Kr14,Go11a}, is conceptually different from the
traditional way~\cite{Pa97,It08,It10,Bau10,Bau11,Ga04,Ga03}.
This can be seen immediately by confronting   expression \rf{13}
 with~\cite[Eq. (7)]{Bau10}.
 Instead
of comparing different bare proton contributions $\Gamma^0_p$, $\Gamma^0_n$, $\Gamma^0_{2},\cdots$,
which are not directly measured but are extracted by the experimentalist from the data,
 we compare the total decay rate $\Gamma_{p}$ and the corresponding spectra,
 which include all of the protons that  come from the NMWD. Different
extraction procedures are not unambiguous, as we have discussed in the first version of the
present work \cite{Co12} regarding the derivation of $\Gamma^0_{2}$ by both KEK \cite{Ki09}, and FINUDA
\cite{Ag10}. More details on the second extraction procedure are given
in the next section.

 Fig. \ref{Fig3} is just the replica of Fig. \ref{Fig2}  except for the numbers of protons, with
$\Delta {\rm N}^{th}_p(E_p)$ evaluated from \rf{14}.
For reasons of completeness, in the last figure we show the spectra of all measured hypernuclei.
 Obviously,   the two figures lead to the same conclusions.
The Gaussian-function fits
of each proton spectrum from $80$ MeV onwards, that were performed in
Ref.~\cite{Ag10} and that will be discussed in the next section, are displayed also in the last figure.


\section{Extraction of branching ratio $\Gamma^0_{2}/\Gamma^0_{NM}$ from the data}
\label{Sec5}

The determination of the  branching ratio $\Gamma^0_{2}/\Gamma^0_{NM}$ by FINUDA~\cite{Ag10} is based
on the  partition of the total number of detected protons ${\rm N}_p$ into  low, and high energy
regions populated, respectively, by  ${\rm N}_p^<\equiv {\rm N}_p(E_p<{E_{\mbox{\tiny part}}})$, and
 ${\rm N}_p^>\equiv {\rm N}_p(E_p>{E_{\mbox{\tiny part}}})$ protons relative to
  the partition energy $E_{\mbox{\tiny part}}$. After assuming that all 2N-NM protons are contained
  within ${\rm N}_p^<$, they define  the ratio
\ber
R&\equiv&\frac{{\rm N}_p^<}{{\rm N}_p}
=\frac{{\rm N}^{0<}_p+{\rm N}^{0}_2 +N_{FSI}^<}{{\rm N}^{0}_p+{\rm N}^{0}_2 +N_{FSI}}
=\frac{\Gamma_p^{0<}/\Gamma_p^0+\Gamma^0_2/\Gamma^0_p+N_{FSI}^</{\rm N}_p^0}
{1+\Gamma^0_2/\Gamma^0_p+N_{FSI}/{\rm N}_p^0},
\label{20}
\eer
where ${\rm N}^{0}_p$ and ${\rm N}^{0}_2$ are, respectively, the numbers of protons induced by the
  1N-NM, and 2N-NM decays, and ${\rm N}_{FSI}={\rm N}^<_{FSI}+{\rm N}^>_{FSI}$ is the total number of particles produced by the FSIs.

 The next steps done  in  Ref.~\cite{Ag10}
 are not supported by sufficiently firm physical arguments. Namely, it is assumed: i)
  that the  proton spectra from $80$ MeV onwards   are due entirely  to protons coming
from the $\Lambda p\go np$ reaction, ii) that they can be fit by Gaussian curves
shown in  Fig. \ref{Fig3}, and
iii) that the maxima of these curves correspond to the partition energies $E_{\mbox{\tiny part}}$.
 All of this yields
\ber
R&\equiv&\frac{{\rm N}_p^<}{{\rm N}_p}
=\frac{0.5+\Gamma^0_2/\Gamma^0_p+N_{FSI}^</{\rm N}_p^0}
{1+\Gamma^0_2/\Gamma^0_p+N_{FSI}/{\rm N}_p^0},
\label{21}
\eer
since the Gaussian curves are bell shaped, satisfying always  the condition
$\Gamma_p^{0<}\equiv \Gamma^0_p(E_p<{E_{\mbox{\tiny max}}})=
 \Gamma_p^{0>}\equiv \Gamma^0_p(E_p>{E_{\mbox{\tiny max}}})$. The resulting partition energies
 $E^{\mbox{\tiny FINUDA}}_{\mbox{\tiny max}}$ are listed in the second column of Table  \ref{T4}.
\begin{table}[htpb]
\caption{Mean values of FINUDA Gaussian fits~\cite{Ag10}(in units of MeV) are confronted with
 the maxima of proton spectra  with the theoretical maxima (third column), as well as with the energies
 ${ E}^{\mbox{\tiny th}}_{\mbox{\tiny even}}$ for which the proton strength is evenly distributed,
 \ie $\Gamma_p^{0>}\equiv \Gamma_p^{0<}$  (fourth column).
} \label{T4}
\bigskip
\begin{center}
\begin{tabular}{cccc}
\hline\noalign{\smallskip}
Hypernucleus    &${ E}^{\mbox{\tiny FINUDA}}_{\mbox{\tiny max}}$
&${ E}^{\mbox{\tiny th}}_{\mbox{\tiny max}}$&${ E}^{\mbox{\tiny th}}_{\mbox{\tiny even}}$\\
\noalign{\smallskip}
\hline\noalign{\smallskip}
$_\Lambda^{5}$He &$68.5\pm4.1$&$77.5$&$75.0$\\
$_\Lambda^{7}$Li &$76.7\pm5.2$&$73.5$&$72.0$\\
$_\Lambda^{9}$Be &$78.2\pm6.2$&$69.0$&$69.0$\\
$_\Lambda^{11}$B &$75.1\pm5.0$&$69.0$&$70.5$\\
$_\Lambda^{12}$C &$80.2\pm 2.1$&$67.5$&$70.5$\\
$_\Lambda^{13}$C &$83.9\pm12.8$&$67.5$&$70.5$\\
$_\Lambda^{15}$N &$88.1\pm6.2$&$61.5$&$66.0$\\
$_\Lambda^{16}$O &$93.1\pm6.2$&$61.5$&$66.0$\\
\noalign{\smallskip}
\hline
\end{tabular}
\end{center}
\end{table}

 Next, FINUDA approximated \rf{21}
by a linear function of the mass number  $A$, \ie
\be
R(A)=a+bA,
\label{22}\ee
where
\be
a=\frac{0.5+\Gamma^0_2/\Gamma^0_p}
{1+\Gamma^0_2/\Gamma^0_p},
\label{23}\ee
 does not depend on $A$. Finally,   a $\chi^{2}$ fit  for $R(A)$ was done for the energies ${ E}^{\mbox{\tiny FINUDA}}_{\mbox{\tiny max}}$
to  obtain the values of   $a$ and  $b$
that are shown in  row A of Table \ref{T5}, together with the
resulting  $\Gamma^0_{2}/\Gamma^0_{p}$, and $\Gamma^0_{2}/\Gamma^0_{NM}$
derived from
\be
 \frac{\Gamma^0_{2}}{\Gamma^0_{p}}=\frac{a-0.5}{1-a},~~
  \frac{\Gamma^0_{2}}{\Gamma^0_{\rm NM}}
  =\frac{a-0.5}{(1-a)\Gamma^0_n/\Gamma^0_p+0.5},
\label{24}
\end{equation}
for  the experimental value $\Gamma^0_n/\Gamma^0_p=0.48 \pm 0.08$, measured by KEK~\cite{Bh07}.

The FINUDA  procedure  to extract the value of $\Gamma^0_{2}$ from  a
series of kinetic energy spectra, by
 separating them into low and  high energy regions, looks physically sound.
However, we shall  soon see that it is very sensitive to the
separation procedure.
On the other hand,  in the fitting of the spectra with Gaussian curves,
FINUDA  implicitly assumes
 the absence of  $N_{FSI}^>$, which is not only unrealistic, but also not necessary.

Before separating the spectra, we compare the calculated
 energy locations $E^{\mbox{\tiny th}}_{\mbox{\tiny max}}$ of the spectra maxima
  with the maxima of the FINUDA Gaussian fits $E^{\mbox{\tiny FINUDA}}_{\mbox{\tiny max}}$.
From Figs. \ref{2} and \ref{3}, one immediately notices sizeable differences.
Moreover, from Table \ref{T4}, one sees that,
while the maxima $E^{\mbox{\tiny FINUDA}}_{\mbox{\tiny max}}$  increase from
$68.5$ MeV to $93.1$  MeV in going from $_\Lambda^{5}$He to $_\Lambda^{16}$O, the energies
$E^{\mbox{\tiny th}}_{\mbox{\tiny max}}$ decrease from $77.6$ MeV to $61.5$ MeV.
On the other hand, while  the Gaussian curves are bell shaped,
 the theoretical 1N-NM spectra deviate significantly from a symmetrical shape.
More precisely, the calculated $\Gamma_p^{0<}$ is greater than $\Gamma_p^{0>}$ in
$_\Lambda^{5}$He, $_\Lambda^{7}$Li, and
$_\Lambda^{9}$Be, and smaller for the remaining hypernuclei.

\begin{figure}[h]
\begin{center}
\resizebox{1.0\textwidth}{!}{
\begin{tabular}{cc}
\includegraphics[width=0.5\linewidth,clip=]{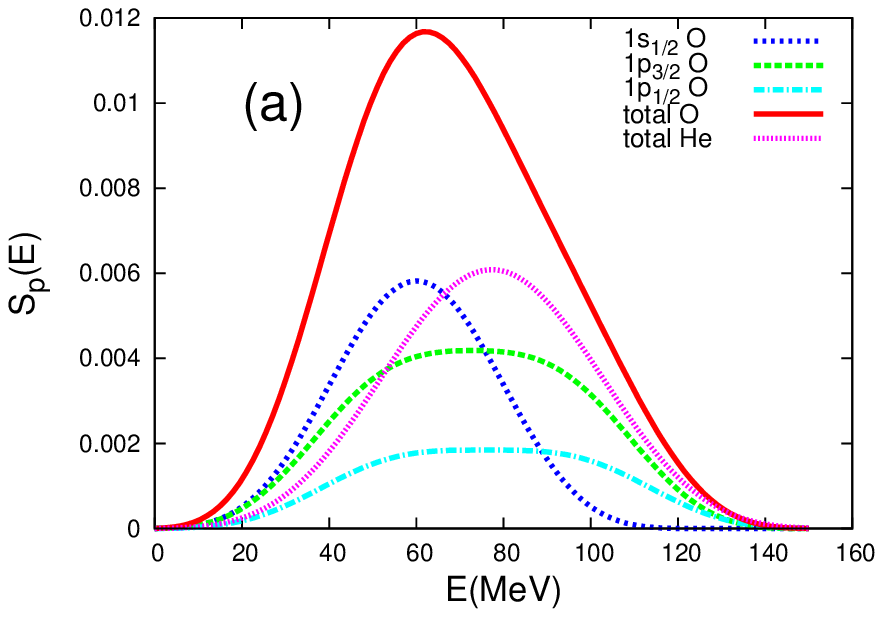}&
\includegraphics[width=0.5\linewidth,clip=]{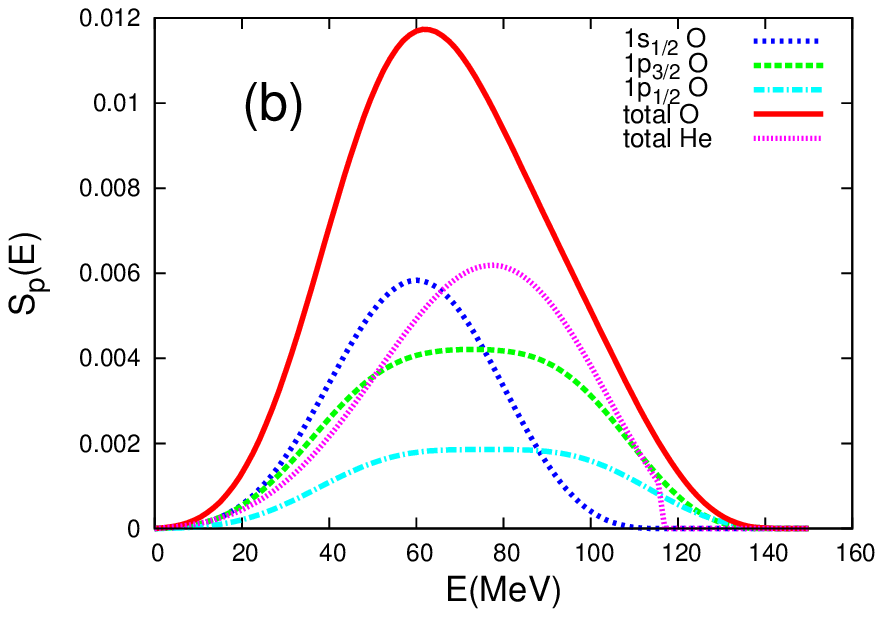}
\end{tabular}}
\caption{\label{Fig4} (Color online) $^{5}_\Lambda$He and  $^{16}_\Lambda$O spectra:
(a) without recoil, and  (b) with recoil,  evaluated within the IPSM for the P2
OME transition potential. Contributions of different  orbitals $s_{1/2}$, $p_{3/2}$,
and $p_{1/2}$ to the total $^{16}_\Lambda$O spectra are also displayed.}
\end{center}
\end{figure}

The reason for this can be  understood from an  inspection of   Fig.
\ref{Fig4}, where the spectra of $_\Lambda^{5}$He and
$_\Lambda^{16}$O without recoil (a)
 and with recoil (b) are
displayed.
First, as expected, the recoil effect is sizeable  in $_\Lambda^{5}$He where
only the  orbital $s_{1/2}$ contributes.
Second, the $_\Lambda^{5}$He spectrum  does not have the   symmetric bell shape, mainly because
 the single kinetic energy reaches its maximum value rather abruptly at $\sim 115$ MeV
 due to  the recoil factor $(A-2)/(A-1)=3/4$ in Eq. \rf{18} for $Q^{s_{1/2}}_p$;
this effect, however, does not modify the value of
${E}_{\mbox{\tiny max}}=77.6$ MeV, but causes $\Gamma_p^{0<}$ to be appreciably  larger than $\Gamma_p^{0>}$
 ($\Gamma_p^{0<}\cong 0.55~\Gamma^0_p$).
Third, in the case of $_\Lambda^{16}$O, three partial waves
($s_{1/2}$, $p_{3/2}$, and $p_{1/2}$) contribute with different heights and
widths, the convolution of which is a nonsymmetric proton spectrum
with $\Gamma_p^{^0<}=0.43~\Gamma^0_p$; here the energy ${E}_{\mbox{\tiny max}}$ becomes
significantly smaller because the  energy
$\Delta^{s_{1/2}}_{p}$, given by \rf{15}, is  $\sim 25$ MeV
smaller in  $_\Lambda^{16}$O than in  $_\Lambda^{5}$He.
 Briefly, as
  the value of $A $ increases, the average value of the binding energies
 $\varepsilon_{\Lambda}$ and
$\varepsilon^j_{N}$ turns out to be larger  (see Ref.~\cite[Fig. 11]{Ja73}),
which  makes the energy position of the maximum ${E}_{\mbox{\tiny max}}^{\mbox{\tiny th}}$
for the $1N$-NM proton kinetic energy  spectrum become increasingly  smaller.
The numerical results for ${ E}_{\mbox{\tiny max}}^{\mbox{\tiny th}}$ are shown in Table \ref{T4}, where they are compared with
the peaks of the FINUDA Gaussian fits ${ E}_{\mbox{\tiny max}}^{\mbox{\tiny FINUDA}}$~\cite{Ag10}.
Note  that ${ E}_{\mbox{\tiny max}}^{\mbox{\tiny th}}$ decreases with the mass number while
${ E}_{\mbox{\tiny max}}^{\mbox{\tiny FINUDA}}$ increases. Thus,
they differ from one another quite significantly and the difference
increases  from   $9$ MeV in  $_\Lambda^{5}$He  up to 32 MeV in
$_\Lambda^{16}$O.

The energies ${ E}_{\mbox{\tiny max}}$ are closely related to the liberated energies, \ie to the ${Q}^j_p$-values
since the latter should, in principle, also decrease when  the $\Delta^j_{p}$  decrease. However,
for light hypernuclei, this decrease is
largely offset by the recoil effect, as shown by Eq. \rf{18}. The final results,
 displayed in Figs. \ref{Fig2} and \ref{Fig3}, demonstrate
that the Q-values  are
roughly constant and within the energy interval of $\sim 115-135$ MeV,
which is consistent with the data within  experimental errors.
\footnote{
One should also mention that it is assumed here that the residual nucleus is
emitted in the ground state, and that consideration of  excitation energies could further
diminish the Q-values.
After finishing this work, we learned  that Bufalino~\cite{Bu13}, one of the coauthors of Ref.
 ~\cite{Ag10}, has proposed evaluating partition energies as half the Q-values
in the $150-166$ MeV range.
 In this way, she obtains good agreement with ${ E}^{\mbox{\tiny FINUDA}}_{\mbox{\tiny max}}$
 for $A$ from $5$ to $9$, whereas for $A = 13,15$,
 and $16$ there is a $2\sigma$ discrepancy. Note that the last range for
 the Q-values ​​implies unrealistically  small proton separation energies.}

\begin{table}[pt]
\centering \caption{Results for  the $\chi^{2}$ parameters $a$ and
$b$, and the corresponding ratios ${\Gamma^0_{2}}/{\Gamma^0_{p}}$, and
$\Gamma^0_{2}/\Gamma^0_{\rm NM}$ for FINUDA data~\cite{Ag10},
and different partition energies for ${\rm N}_p^<$, and ${\rm N}_p^>$:
 A) derived in  Ref.~\cite{Ag10} with ${ E}^{\mbox{\tiny FINUDA}}_{\mbox{\tiny max}}$,
  B) and C) obtained here with ${ E}^{\mbox{\tiny th}}_{\mbox{\tiny max}}$, and
 ${ E}^{\mbox{\tiny th}}_{\mbox{\tiny even}}$, respectively.}
 \label{T5}
\bigskip
\small\addtolength{\tabcolsep}{-3pt}
\begin{tabular}{ccccc}
\hline \noalign{\smallskip}
case&$a$&$b$&${\Gamma^0_{2}}/{\Gamma^0_{p}}$&$\Gamma^0_{2}/\Gamma^0_{\rm NM}$ \\
\noalign{\smallskip}
\hline
\noalign{\smallskip}
A&$0.654\pm0.138$&$  0.009\pm  0.013 $&$ 0.43\pm 0.25$&$ 0.24\pm 0.10$\\
B&$0.711\pm  0.101$&$ -0.010\pm  0.008$&$   0.73\pm   0.61$&$   0.33\pm   0.10$\\
C&$  0.656\pm  0.101$&$   -0.004\pm   0.008$&$   0.45\pm   0.43$&$   0.23\pm   0.09$\\
\noalign{\smallskip}
\hline
\end{tabular}
\end{table}

\begin{figure}[h]
\begin{center}
\psfig{file=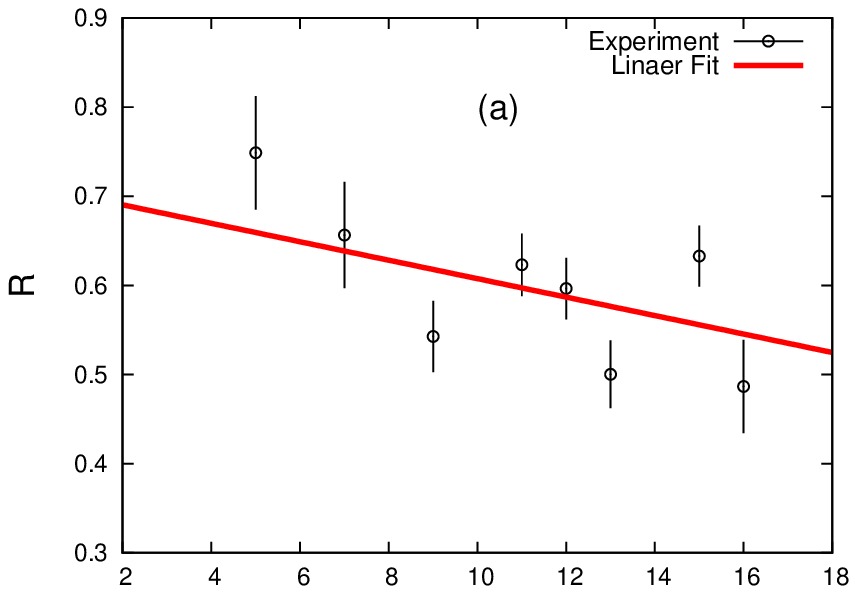,width=6cm} \\
\psfig{file=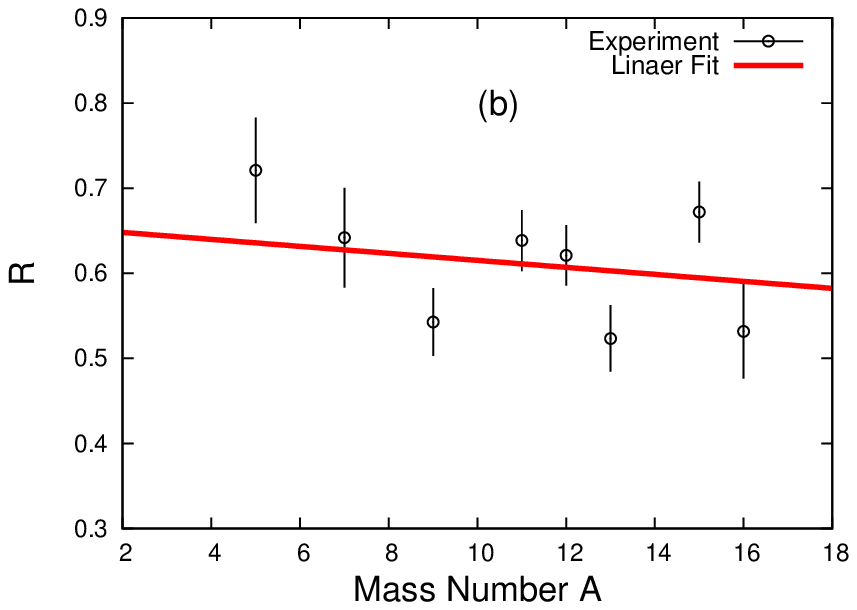,width=6cm}
\caption{\label{Fig5} (Color online) Ratio $R$, given by \rf{11} as a function of
$A$ for  the partition energies: (a) ${ E}_{\mbox{\tiny max}}^{\mbox{\tiny th}}$,
and (b) ${ E}^{\mbox{\tiny th}}_{\mbox{\tiny even}}$.}
\end{center}
\end{figure}

In view of the differences between the  FINUDA spectra and those
calculated here, one is  immediately tempted
to repeat the above analysis using the energies
${ E}^{\mbox{\tiny th}}_{\mbox{\tiny max}}$ instead of
${ E}^{\mbox{\tiny FINUDA}}_{\mbox{\tiny max}}$. This was done
and the results for $R$ are shown in the left panel of Figure \ref{Fig5},
along with the corresponding linear $\chi^2$ fit. The
large difference between the two sets of maxima
gives rise to large differences between the values of the
ratios $R$ in  Ref.~\cite[Fig. 2]{Ag10}  and those derived here.
The  parameters $a$ and
$b$, and the  ratios ${\Gamma^0_{2}}/{\Gamma^0_{p}}$, and
$\Gamma^0_{2}/\Gamma^0_{\rm NM}$ obtained in this way are listed in  row B of Table \ref{T5}.
The value of $a$ is not very different from the previous case but,
as expected,  $b$ is now negative, and the resulting
 $\Gamma^0_{2}$ is significantly different due to its strong sensitivity
on $a$ in \rf{24}.

One must not forget here that, while  the number of protons can be partitioned in many different ways,
obtaining different results for the ratio defined in \rf{20}, the relations from  \rf{21} on
are only valid when
 the condition $\Gamma_p^{0<}=\Gamma_p^{0>}=\Gamma^0_p/2$
 is fulfilled, which does not occur for ${ E}_{\mbox{\tiny max}}^{\mbox{\tiny th}}$.
There is, however, always an energy
${ E}^{\mbox{\tiny th}}_{\mbox{\tiny even}}$ for which  this condition is fulfilled, and which
are listed in the last column of Table \ref{T4}.  The values of new $R$, displayed in the lower panel of
 Fig. \ref{Fig5}, are not very different from the previous values shown in the same figure.
However, the value of the parameter $a$
and, consequently, the ratio ${\Gamma^0_{2}}/{\Gamma^0_{p}}$, is quite different now
resembling those obtained by FINUDA, as can be seen from  row C in Table \ref{T5}.
This agreement is somewhat surprising and we can not draw any conclusion from it. One has just
 learned  that: (i) a relatively small modification of the  partition energies (from
 ${ E}_{\mbox{\tiny max}}^{\mbox{\tiny th}}$ to ${ E}^{\mbox{\tiny th}}_{\mbox{\tiny even}}$) can lead to  significantly different results
for ${\Gamma^0_{2}}/{\Gamma^0_{p}}$, and (ii) a relatively sizeable  modification
 of the  partition energies (from ${ E}_{\mbox{\tiny max}}^{\mbox{\tiny FINUDA}}$ to ${ E}^{\mbox{\tiny th}}_{\mbox{\tiny even}}$)
 can lead to  quite similar results
for ${\Gamma_{2}^0}/{\Gamma^0_{p}}$.

It would be valuable  to find the physical meaning of the fitting parameter $b$,
 which is different in the three cases discussed above. In this regard,
 how to arrive  at \rf{22} from \rf{21} is not a trivial issue.
One possibility is to neglect the last term in the denominator of
\rf{21}, arguing, as was done in Ref. \cite{Ag10}, that the FSIs
tend to remove protons from the high energy part of the spectrum
(${\rm N}_{FSI}^><0$) while filling the low energy region (${\rm
N}_{FSI}^<>0$), with the net result that ${\rm N}_{FSI}={\rm
N}_{FSI}^<+{\rm N}_{FSI}^>\cong 0$.
Therefore
\brn
R\cong a+\frac{{\rm N}_{FSI}^<}{(1+\Gamma^0_2/\Gamma^0_p){\rm
N}_p^0},
\ern
which, when compared with \rf{22},
yields
\be {{\rm N}_{FSI}^<}\cong b{A}{\rm N}_p^0 \left(1+\frac{\Gamma_2^0}{\Gamma_p^0}\right)\sim b{A}{\rm N}_p^0,
\label{25}\ee
 since the factor
${1+\Gamma^0_2/\Gamma^0_p}< 1.5$ is unsubstantial for a qualitative
discussion.
At first glance, the last equation  appears reasonable
because the effect of the FSIs should increase with $A$.
But, since $b=0.009\pm 0.013$, it turns out that
 ${\rm N}_{FSI}^</{\rm N}_p^0\sim 0.01A$.
Such a small amount of FSIs looks  unrealistic.
It is even more difficult to interpret physically the negative values of  $b$ that we  obtain in cases B and C.
Evidently, the fact that the IPSM is unable to reproduce the low-energy  spectra in no way affects the previous discussion of ${\Gamma^0_{2}}/{\Gamma^0_{p}}$.

A different  derivation of the $2N$ branching ratio
 has been done at FINUDA quite recently~\cite{Ag11},
based on the analysis of the ($\pi^-, p, n$) triple coincidence events, and
a $\chi^2$ fit similar to \rf{22}. The new result, $\Gamma^0_{2}/\Gamma^0_{\rm NM}=021\pm 0.10$,
 is consistent within the errors  with the previously
 obtained value~\cite{Ag10}, as well as with our result.
Only the sum of   events from all hypernuclear species are exhibited in this work,
without presenting data for individual   proton spectra, which would allow us to do
  a reanalysis similar to that done above.


\section{Summary and Conclusions}
\label{Sec6}
Proton kinetic energy spectra of ${\mathrm{^{5}_{\Lambda}He}}$,
${\mathrm{^{7}_{\Lambda}Li}}$, ${\mathrm{^{9}_{\Lambda}Be}}$, ${\mathrm{^{11}_{\Lambda}B}}$, ${\mathrm{^{12}_{\Lambda}C}}$,
${\mathrm{^{13}_{\Lambda}C}}$, ${\mathrm{^{15}_{\Lambda}N}}$ and ${\mathrm{^{16}_{\Lambda}O}}$,
 measured by FINUDA a few years ago~\cite{Ag10}, were
evaluated theoretically for the first time.
We conclude that, in all the cases, the magnitudes of the  spectra strongly depend on the parameterization that is used for  the transition potential, while  their shapes are very similar and independent of the transition mechanism.
This statement, like all statements made in this work, do not depend at all on the inclusion or non-inclusion of the FSIs and $2N$-NM channel in the nuclear model.

It is explained in detail in Sec. \ref{2} that our method for  comparing the theory with experiment
  is radically
different from the  procedure followed by other researchers.
In particular:
\bit
\item  The equations \rf{4} have never been used so far by any other group.
These relations  show that, to evaluate  the decay rates, it is essential to
know the number of produced hypernuclei ${\rm N}_W$, which is not available in the literature.

\item We focus our attention on  measured transition probabilities $\Gamma_N$, and
$\Gamma_{nN}$, instead of comparing  bare quantities $\Gamma^0_p$,
$\Gamma^0_n$, $\Gamma^0_{np}$, \etc,  which are extracted from the experiments by making assumptions and  approximations that are often questionable.

\item The difference with other studies can be immediately seen 
by facing our  Eq. \rf{14} with ~\cite[Eq. (7)]{Bau10}.

\item But, what is really important is that the method proposed here imposes
  more constraint in comparing theory with data, allowing us to examine  more clearly the
  decay mechanism  regardless of the importance of FSIs and the 2N-NM channel.
\eit
Despite the lack of direct information about the ${\rm N}_W$-values,
we have been able to estimate  these observables  for ${\mathrm{^{5}_{\Lambda}He}}$,
${\mathrm{^{7}_{\Lambda}Li}}$, and ${\mathrm{^{12}_{\Lambda}C}}$ hypernuclei
 from the ratios $R_p$ presented in Ref.~\cite{Ag08}.
(The physical meaning of this ratio is also clarified.)
In this way,  we obtain in Sec. \ref{3}
some very useful information on the transition potential.
In fact, from Figs. \ref{Fig2} and \ref{Fig3} and Table  \ref{T1}, we conclude that:
\bit
\item The IPSM  reproduces correctly
    the shapes of all  proton kinetic energy spectra $S_p(E_p)$ at medium and high energies
    ($E_p\gsim 40$ MeV), including the $Q$-value, which is around $115-135$  MeV.
The latter could indicate that the residual nucleus $^{A-2}(Z-1)$ is emitted mainly in
the ground state.

\item This simple model also reproduces fairly well the magnitudes of
the spectra at these  energies
 when the soft $\pi+K$ potential is used to describe the NMWD.
In no way do we claim that this is the "true" physics, but
we believe that it might be worth pursuing  this direction,
especially considering that the model is able to explain
satisfactorily the NM decay rates $\Gamma_p$ and  $\Gamma_n$ of the s-shell hypernuclei~\cite{Bau09,Kr14}.

\item The measured transition rates, $\Gamma_p$, the  same as the
corresponding  spectral densities $S_p(E_p)$ for $E_p\gsim 40$ MeV, are significantly overestimated
by the present theoretical calculations  when the standard parametrization for the
transition potential  is used.
In doing this comparison, one should keep in mind
that, while the calculations  refer to the $1N$-NM decay mode only, the measured rates
include also the $2N$-NM decay channel and the effects of FSIs, and therefore
the latter should always be larger.
 This happens only for the soft $\pi+ K$ exchange potential.

\item It is  difficult to reconcile the FINUDA data with the theory based on the
 $\pi+2\pi/\rho+2\pi/\sigma+\omega+K+\rho\pi/a_1+\sigma\pi/a_1$ exchange potential~\cite{It10}.
\eit

We strongly believe that, in recent  theoretical calculations~\cite{Bau10,Bau11}, which include both the $2N$-NM decay channel, and the FSIs, they would have arrived at very similar conclusions if the comparison with experimental data
have been made ​​in the manner proposed here.

Since the calculated  spectra only involve the $1N$-NM channel,
 the differences between them and the experimental spectra,
 both with respect to the magnitude and in relation to the energy distribution, can indicate
 which other degrees of freedom are important ($2N$-NM channel, FSIs, \etc~).
Our plan of action is to add them to $1N$-NM, when necessary,
in order to successfully reproduce the experiments.

With regard to the  FINUDA method~\cite{Ag10} to determine  the $2N$-NM decay rate,
 based on the assumption that the $1N$-NM strength  of the  kinetic proton spectra
  is equally distributed in the low and high energies, we conclude that:
\bit
\item The proposed method is very sensitive to the energies that separate these two regions,
and these energies can not be determined experimentally.

\item The separation  done  in  Ref.~\cite{Ag10} is not supported by any firm physical argument.

\item  It is necessary to resort to theoretical models to establish the partition energies;
the IPSM is very suitable for this purpose.

\item Both theoretical sets of partition energies (${ E}^{\mbox{\tiny th}}_{\mbox{\tiny max}}$,
 ${ E}^{\mbox{\tiny th}}_{\mbox{\tiny even}}$) differ from the FINUDA result
 (${ E}^{\mbox{\tiny FINUDA}}_{\mbox{\tiny max}}$),
not only in magnitudes but also with regards to the mass-number dependence:
 ${ E}^{\mbox{\tiny th}}_{\mbox{\tiny max}}$ and  ${ E}^{\mbox{\tiny th}}_{\mbox{\tiny even}}$
 decrease with $A$, because  the experimental single-particle energies  increase;
 meanwhile,  no reasonable  explanation  exists  for the  opposite behavior of
 ${ E}^{\mbox{\tiny FINUDA}}_{\mbox{\tiny max}}$.

\item In spite of the above mentioned differences, all three sets of partition energies yield similar results
for the parameter $a$  and therefore for the ratio $\Gamma^0_2/\Gamma^0_p$.
This indicates that the behavior with $A$ of the partition energies does not play a crucial role,
and is consistent with a recent proposal to approximate them  by a rather constant value of $\lsim 80$ MeV
~\cite{Bu13}.

\item Physical interpretation of the   FINUDA parameter $b$
is  a  point at issue, not only for being very small, but also because it
is negative in our analysis, as well as  in the a new study~\cite{Ag11} of
 the contribution of the $2N$-NM channel employing the same method.
Although suffering from large errors, its small value inevitably leads to the conclusion that the FSIs are very small, and that, therefore, the low-energy proton spectra dominantly comes from the $2N$-NM decay.
It is very hard to understand this fact, and the only alternative possibility
is that the basic approach \rf{22} was incorrect.
\eit
Final remarks:
\bnu
\item  As stated in the beginning, our purpose was not to reproduce  the experimental data,
but to discover out what the proton kinetic energy spectra   can tell us about the weak
hypernuclear interaction.
 To account for the  low energy data of the kinetic energy spectra,
 it is imperative to consider the FSIs.
For the time being, we are working on this issue by employing an improved version of the CRISP
internuclear cascade model~\cite{Ba13} used previously to describe
  ${\mathrm{^{12}_{\Lambda}C}}$~\cite{Go11,Go11a}.

  Moreover, a complete theoretical
  description must also include a judicious estimate of the effect of the
  $2N$-NM decay channel. So far, this has been done only in the context of
  FGM~\cite{Bau10,Bau11},   and it would be interesting to know
   what the SM can tell us about  this process.
  In fact, for quite some time, we have been  involved in the
  development of a corresponding theoretical formalism and  numerical codes~\cite{Ba13a}.

\item To study the FSIs and the $2N$-NM decay, it
 is indispensable to understand  first the $1N$-NM-decay dynamics.
In our opinion,  the SM could be a very useful tool to achieve this goal.
 For instance, the SM spectra are
 the main ingredients for establishing the initial conditions for the
 FSIs within the many-body multi-collision
Monte Carlo cascade scheme~\cite{Go11,Go11a}.
On the other hand, from Fig. \ref{Fig1},
it is self-evident that
the diagram a) plays the principal
role within the diagram b) representing the $2N$-NM decay mode.

\item  New experimental developments  will be
very welcome, such as:
i) Angular correlation of $np$ and $nn$  pairs to determine the
$\Gamma_{np}$ and $\Gamma_{nn}$ rates, which have been  measured so far only by
KEK~\cite{Ki06} in $_\Lambda^{12}$C, and
ii) Triple $(p,n,n)$, and $(p,p,n)$ coincidence detections
 for direct measurement of $\Gamma_{pn}$, and $\Gamma_{pp}$,
 as suggested previously~\cite{Co12}. The first steps in this direction seems to be  given
 recently in Ref.~\cite{Bu13}.

\item To complete  Figs. \ref{2} and \ref{3}, we need the  ${\rm N}_W$-values for ${\mathrm{^{9}_{\Lambda}Be}}$, ${\mathrm{^{11}_{\Lambda}B}}$,
${\mathrm{^{13}_{\Lambda}C}}$, ${\mathrm{^{15}_{\Lambda}N}}$ and ${\mathrm{^{16}_{\Lambda}O}}$  hypernuclei.
  Hopefully, these numbers will soon be available for public use.​
Needless to point out that, otherwise, the  proton kinetic energy spectra measured by FINUDA~\cite{Ag10} are of little  use to study the
NMWD dynamics. They only can be exploited to discuss the decay kinematics through the analysis of the FSIs and the $2N$-NM decay mode.
\enu


\section*{Acknowledgements}
FK is supported by  by  Argentinean agencies CONICET (PIP 0377) and FONCYT (PICT-2010-2680),
as well as by the Brazilian agency FAPESP (CONTRACT 2013/01790-5).
We are very grateful to Dr.
Gianni Garbarino and Dr. Airton Deppman for very enlightening discussions and to
Dr. Wayne Allan Seale for the careful and critical reading of the manuscript.





\begin{thebibliography}{0}
\bibitem{Al02} W.M. Alberico, G. Garbarino, Phys. Rep. \textbf{369},  (2002) 1.
\bibitem{St94} V. G. J. Stoks, R. A. M. Klomp, C. P. F. Terheggen, and J. J. de Swart,  Phys. Rev. C \textbf{49},  (1994) 2950.
\bibitem{Du96} J. F. Dubach, G. B. Feldman, B. R. Holstein, L. de la Torre,
Ann. Phys. (N.Y.) \textbf{249}, (1996) 146.
\bibitem{Pa97} A. Parre\~{n}o, A. Ramos, and C. Bennhold, Phys. Rev. C \textbf{56}, (1997) 339.
\bibitem{Boc99} R. Bockmann, C. Hanhart, O. Krehl, S. Krewald, and J. Speth, Phys. Rev. \textbf{C 60}, (1999) 055212.
\bibitem{Nak12} K. Nakayama, privite comunication (2012).
\bibitem{Ryc94} J. Ryckebusch, M. Vanderhaeghen, L. Machenil, and
M. Waroquier, Nucl. Phys. A \textbf{568}, (1994) 828.
\bibitem{Bau09} E. Bauer, A.P. Gale\~ao, M. Hussein, F. Krmpoti\'c, and  J.D. Parker,
 Phys. Lett. B \textbf{674}, (2009) 103.
\bibitem{Kr14}  F. Krmpoti\'c, Few-Body Syst  \textbf{55}, (2014) 219.
\bibitem{Kr10a} F. Krmpoti\'c, Phys. Rev.  C \textbf{82},  (2010) 055204.
\bibitem{Pa02} A. Parre\~{n}o, A. Ramos,  Phys. Rev. C \textbf{65},  (2001) 015204.
\bibitem{Sa00} K. Sasaki, T. Inoue, and M. Oka,  Nucl.Phys. A \textbf{669}, (2000) 331; Erratum-ibid. A \textbf{678}, (2000) 455.
\bibitem{Sa02} K. Sasaki, T. Inoue, and M. Oka, Nucl. Phys. A \textbf{707}, (2002) 477.
 \bibitem{Pa04} A. Parre\~no, C Bennhold, and B. R. Holstein, Phys.
Rev. C \textbf{70}, (2004) 051601.
\bibitem{Ch07} C. Chumillas, G. Garbarino, A. Parre\~{n}o, and A. Ramos,
Phys. Lett.  B \textbf{657}, (2007) 180.
\bibitem{It08} K. Itonaga,  T. Motoba,  T. Ueda, and Th.A. Rijken, Phys. Rev. C \textbf{77}, (2008)  044605.
\bibitem{It10} K. Itonaga,  T. Motoba,  Prog. Theor. Phys. Suppl. \textbf{185}, (2010)  252.
\bibitem{Ru56} M. Ruderman and R. Karplus, Phys. Rev. \textbf{102}, (1956) 247.
\bibitem{Al91} W.M. Alberico, A. De Pace, M. Ericson, and A. Molinar, Phys. Lett. B \textbf{256}, (1991) 134.
\bibitem{Ra94} A. Ramos,  E. Oset and L.L. Salcedo, Phys. Rev.  C \textbf{50}, (1994) 2314.
\bibitem{Ki09} M. J. Kim,~\etal, Phys. Rev.  Lett. \textbf{103}, (2009) 182502 .
\bibitem{Ag08} M. Agnello \etal, Nucl. Phys. A \textbf{804}, (2008) 151.
\bibitem{Ag09} M. Agnello \etal, Phys. Lett. B \textbf{681}, (2009) 139.
\bibitem{Ag10} M. Agnello \etal, Phys. Lett. B \textbf{685}, (2010) 247.
\bibitem{Bau09a} E. Bauer,  and G. Garbarino,  Nucl. Phys. A \textbf{828}, (2009) 29.
\bibitem{Pa07} J. D. Parker \etal, Phys. Rev. C \textbf{76}, (2007)   035501.
\bibitem{Sz91} J.J. Szymanski \etal, Phys. Rev. C \textbf{43}, (1991) 849.
\bibitem{No95} H. Noumi \etal, Phys. Rev. C \textbf{52}, (1995) 2936.
\bibitem{Ki03} J. H. Kim \etal, Phys. Rev. C \textbf{68}, (2003) 065201.
\bibitem{Ok04} S. Okada \etal, Phys. Lett. B \textbf{597}, (2004) 249.
\bibitem{Ok05} S. Okada \etal, Nucl. Phys. A \textbf{752 },  (2005) 196.
\bibitem{Ou05} H. Outa \etal, Nucl. Phys. A \textbf{754 }, (2005) 157c.
\bibitem{Ka06} B. H. Kang \etal, Phys. Rev. Lett. \textbf{96}, (2006) 062301.
\bibitem{Ki06} M. J. Kim \etal, Phys. Lett. B \textbf{641}, (2006) 28.
\bibitem{Bh07} H. Bhang \etal, Eur. Phys. J. A \textbf {33}, (2007) 259.
\bibitem{Bau10} E. Bauer, G. Garbarino, A. Parre\~{n}o, and A. Ramos, Nucl. Phys. A \textbf{836}, (2010) 199.
\bibitem{Bau10a} E. Bauer, and G. Garbarino, Phys. Rev. C \textbf{81},   (2010) 064315.
\bibitem{Bau11} E. Bauer,  and G. Garbarino, Phys. Lett. B \textbf{698}, (2011) 306.
\bibitem{Be06} O. Benhar, Nucl. Phys. B (Proc. Suppl.) \textbf{159}, (2006) 168.
\bibitem{Me13} A. Meucci, C. Giusti, M. Vorabbi, Phys. Rev. D {\bf 88} 013006.
\bibitem{Sc66} H. F.  Schopper , {\it Weak Interactions and Nuclear
$\beta$-decay} (North-Holland Publ. Co., Amsterdam, 1966).
\bibitem{Ra97} A. Ramos, M. J. Vicente-Vacas,
 and E. Oset, Phys. Rev.  \textbf{C 55}, 735 (1997);\textbf{C 66},  039903 (2002)(E).
\bibitem {Go11} I. Gonzalez \etal, J. Phys. Conf. Ser. \textbf{312}, (2011)  022017.
\bibitem {Go11a} I. Gonzalez, C. Barbero, A. Deppman,  S. Duarte, F.
Krmpoti\'c, O. Rodriguez, J. Phys. G: Nucl. Part. Phys. \textbf{38},   (2011) 115105.
\bibitem{CRISP1} A. Deppman, S. B. Duarte, G. Silva,  O. A. P. Tavares, S. Anefalos, J. D. T. Arruda-Neto,
T. Rodrigues J. Phys. G: Nucl. Part. Phys. {\bf 30} 1991.
\bibitem{CRISP2} A. Deppman,  G. Silva,  S. Anefalos,  S. B. Duarte, F. Garcia,
F. H. Hisamoto, O. A. P. Tavares Phys. Rev. C {\bf 73} 064607.
\bibitem{CRISP3} A. Deppman, O. A. P. Tavares , S. B. Duarte, E. C. Oliveira, J. D. T. Arruda-Neto, S. de Pina, V.  Likhachev, O. Rodriguez, J. Mesa, M. Gonçalves Phys. Rev. Lett. {\bf 87} 82701
\bibitem{CRISP4}  O. A. P. Tavares , S. B. Duarte, V. Likhachev, A. Deppman
  J. Phys. G: Nucl. Part. Phys. {\bf 30} 37.
\bibitem{Ga04}  G. Garbarino, A. Parre\~{n}o, and A. Ramos,
Phys. Rev. C \textbf{69}, (2004)   054603.
\bibitem{Bot14} E. Botta,{\it Weak Decay of Hypernuclei (FINUDA, KEK) – Experiment},
 SPHERE School + 22nd Indian-Summer School of Physics
 STRANGENESS NUCLEAR PHYSICS, Rez/Prague, Czech Republic (2010).
\bibitem{He86} D. P. Heddle and L.S . Kisslinger, Phys. Rev. C \textbf{33}, (1986) 608.
\bibitem{Co95}J. Cohen, Prog.  Part. Nucl. Phys. \textbf{25}, (1990) 139.
\bibitem {Ba02} C. Barbero, D. Horvat, F. Krmpoti\'{c}, T. T. S. Kuo,
Z. Naran\v{c}i\'c, and D. Tadi\'{c}, Phys. Rev. C \textbf{66}, (2002) 055209.
\bibitem {Kr03} F. Krmpoti\'c, and  D. Tadi\'c, Braz. J. Phys. \textbf{33}, (2003) 187.
\bibitem{Ba03} C. Barbero, C. De Conti, A. P. Gale\~ao and F. Krmpoti\'c, Nucl.
Phys. A \textbf{726}, (2003) 267.
\bibitem{Ba07} C. Barbero, A. P. Gale\~ao, and
F. Krmpoti\'c, Phys. Rev. C \textbf{76},  (2007) 054321.
\bibitem{Ba08} C. Barbero, A. P. Gale\~ao, M. S. Hussein, and F. Krmpoti\'c,
Phys. Rev. C \textbf{78}, (2008) 044312; Erratum-ibid. 059901(E).
\bibitem{Ba10} E. Bauer, A. P. Gale\~ao, M. S. Hussein and F. Krmpoti\'c,
Nucl. Phys. A \textbf{834}, (2010) 599c.
\bibitem{Kr10} F. Krmpoti\'c,  A. P. Gale\~ao, and M.S. Hussein,
{\it  AIP Conf. Proc.} \textbf{1245},  (2010) 51.
\bibitem{Krm05} F. Krmpoti\'c, A. Samana, and A. Mariano,
                Phys. Rev. C \textbf{71}, (2005) 044319 .
\bibitem{Bau09b} E. Bauer,  Nucl. Phys.  A \textbf{818}, (2009) 174.
\bibitem{TUNL} TUNL Nuclear Data Project, www.tunl.duke.edu/nucldata/ (2012).
\bibitem{Mil07} D.J. Millener, Lecture Notes in Physics \textbf{724} (Springer, 2007) 31.
\bibitem{Ga03} G. Garbarino, A. Parre\~{n}o and A. Ramos,  Phys. Rev. Lett. \textbf{91}, (2003) 112501.
\bibitem{Co12} C. de Conti, A. Deppman, F. Krmpoti\'c, arXiv:1202.4223.
\bibitem{Ja73} G. Jacob and T. A. J. Maris, Rev. Mod. Phys. \textbf{45}, (1973) 6.
\bibitem{Bu13} S. Bufalino,   Nucl. Phys.  A \textbf{914}, (2013) 160.
\bibitem{Ag11} M. Agnello \etal, Phys. Lett. B \textbf{701}, (2011) 556.
\bibitem {Ba13} C. Barbero, A. Deppman,  S. Duarte, F. Krmpoti\'c, in preparation.
\bibitem{Ba13a} C. Barbero, E. Bauer, C. De Conti, A. P. Gale\~ao and F. Krmpoti\'c, in preparation.

\end{thebibliography}
\end{document}